\documentclass[trackchanges]{aastex63}

\usepackage{multirow}
\usepackage{booktabs}

\usepackage{soul}

\received{21 December 2022}
\revised{13 February}
\accepted{10 May 2023}
\submitjournal{PASP}

\shorttitle{NIRISS Overview and In-flight Performance}
\shortauthors{Doyon et al.}
\graphicspath{{./}{figures/}}

\begin{document}

\title{The Near Infrared Imager and Slitless Spectrograph for the  {\it James Webb Space Telescope} - I. Instrument Overview and in-Flight Performance}

\correspondingauthor{René Doyon}
\email{rene.doyon@umontreal.ca}

\author[0000-0001-5485-4675]{Doyon, Ren\'e }
\affiliation{Institut Trottier de recherche sur les exoplan\`etes, D\'epartement de physique, Universit\'e de Montr\'eal}
\affiliation{Observatoire du Mont-M\'egantic, Universit\'e de Montr\'eal, C.P. 6128, Succ. Centre-ville, Montr\'eal, H3C 3J7, Québec, Canada.}

\author[0000-0002-4201-7367]{Chris J. Willott}
\affil{NRC Herzberg, 5071 West Saanich Rd, Victoria, BC V9E 2E7, Canada}

\author{John B. Hutchings}
\affil{NRC Herzberg, 5071 West Saanich Rd, Victoria, BC V9E 2E7, Canada}

\author[0000-0003-1251-4124]{Anand Sivaramakrishnan}
\affiliation{Space Telescope Science Institute, 3700 San Martin Drive, Baltimore, MD 21218, USA}
\affiliation{Astrophysics Department, American Museum of Natural History, 79th Street at Central Park West, New York, NY 10024}
\affiliation{Department of Physics and Astronomy, Johns Hopkins University, 3701 San Martin Drive, Baltimore, MD 21218, USA}

\author[0000-0003-0475-9375]{Lo\"ic Albert}
\affiliation{Institut Trottier de recherche sur les exoplan\`etes, D\'epartement de physique, Universit\'e de Montr\'eal}
\affiliation{Observatoire du Mont-M\'egantic, Universit\'e de Montr\'eal, C.P. 6128, Succ. Centre-ville, Montr\'eal, H3C 3J7, Québec, Canada.}

\author[0000-0002-6780-4252]{David Lafreni\`ere}
\affiliation{Institut Trottier de recherche sur les exoplan\`etes, D\'epartement de physique, Universit\'e de Montr\'eal}

\author[0000-0002-1715-7069]{Neil Rowlands}
\affiliation{Honeywell Aerospace \#100, 303 Terry Fox Drive, Ottawa,  ON  K2K 3J1, Canada} 

\author[0000-0003-3504-1569]{M. Bego\~na Vila}
\affiliation{NASA Goddard Space Flight Center, 8800 Greenbelt Rd, Greenbelt, MD 20771}
\affiliation{KBR Space Engineering Division, 8120 Maple Lawn Blvd, Fulton, MD 20759}

\author{Andr\'e R. Martel}
\affiliation{Space Telescope Science Institute, 3700 San Martin Drive, Baltimore, MD 21218, USA}

\author[0000-0002-5907-3330]{Stephanie LaMassa}
\affil{Space Telescope Science Institute, 3700 San Martin Drive, Baltimore, MD 21218, USA}

\author{David Aldridge}
\affiliation{Honeywell Aerospace \#100, 303 Terry Fox Drive, Ottawa,  ON  K2K 3J1, Canada} 

\author[0000-0003-3506-5667]{\'Etienne Artigau}
\affiliation{Institut Trottier de recherche sur les exoplan\`etes, D\'epartement de physique, Universit\'e de Montr\'eal}
\affiliation{Observatoire du Mont-M\'egantic, Universit\'e de Montr\'eal, C.P. 6128, Succ. Centre-ville, Montr\'eal, H3C 3J7, Québec, Canada.}



\author{Peter Cameron}
\affiliation{Honeywell Aerospace \#100, 303 Terry Fox Drive, Ottawa,  ON  K2K 3J1, Canada} 

\author[0000-0001-7653-0882]{Pierre Chayer}
\affiliation{Space Telescope Science Institute, 3700 San Martin Drive, Baltimore, MD 21218, USA}

\author[0000-0003-4166-4121]{Neil J. Cook} 
\affiliation{Institut Trottier de recherche sur les exoplan\`etes, D\'epartement de physique, Universit\'e de Montr\'eal}

\author[0000-0001-7864-308X]{Rachel A. Cooper}
\affiliation{Space Telescope Science Institute, 3700 San Martin Drive, Baltimore, MD 21218, USA}

\author[0000-0002-7786-0661]{Antoine Darveau-Bernier}
\affiliation{Institut Trottier de recherche sur les exoplan\`etes, D\'epartement de physique, Universit\'e de Montr\'eal}


\author{Jean Dupuis}
\affiliation{Canadian Space Agency, 6767 Route de l'Aéroport, Saint-Hubert, QC J3Y 8Y9, Canada}

\author{Colin Earnshaw}
\affiliation{Honeywell Aerospace \#100, 303 Terry Fox Drive, Ottawa,  ON  K2K 3J1, Canada} 

\author[0000-0001-9513-1449]{N\'estor Espinoza}
\affil{Space Telescope Science Institute, 3700 San Martin Drive, Baltimore, MD 21218, USA}

\author[0000-0002-0201-8306]{Joseph C. Filippazzo}
\affil{Space Telescope Science Institute, 3700 San Martin Drive, Baltimore, MD 21218, USA}

\author[0000-0003-2429-7964]{Alexander W. Fullerton}
\affiliation{Space Telescope Science Institute, 3700 San Martin Drive, Baltimore, MD 21218, USA}

\author{Daniel Gaudreau}
\affiliation{Canadian Space Agency, 6767 Route de l'Aéroport, Saint-Hubert, QC J3Y 8Y9, Canada}

\author{Roman Gawlik}
\affiliation{Honeywell Aerospace \#100, 303 Terry Fox Drive, Ottawa,  ON  K2K 3J1, Canada} 

\author[0000-0002-5728-1427]{Paul Goudfrooij}
\affil{Space Telescope Science Institute, 3700 San Martin Drive, Baltimore, MD 21218, USA}

\author{Craig Haley}
\affiliation{Honeywell Aerospace \#100, 303 Terry Fox Drive, Ottawa,  ON  K2K 3J1, Canada} 

\author[0000-0003-2769-0438]{Jens Kammerer}
\affiliation{Space Telescope Science Institute, 3700 San Martin Drive, Baltimore, MD 21218, USA}

\author{David Kendall}
\affiliation{Canadian Space Agency, 6767 Route de l'Aéroport, Saint-Hubert, QC J3Y 8Y9, Canada}

\author{Scott D. Lambros}
\affiliation{NASA Goddard Space Flight Center, 8800 Greenbelt Rd, Greenbelt, MD 20771}

\author{Luminita Ilinca Ignat}
\affiliation{Canadian Space Agency, 6767 Route de l'Aéroport, Saint-Hubert, QC J3Y 8Y9, Canada}

\author{Michael Maszkiewicz}
\affiliation{Canadian Space Agency, 6767 Route de l'Aéroport, Saint-Hubert, QC J3Y 8Y9, Canada}

\author{Ashley McColgan}
\affiliation{Honeywell Aerospace \#100, 303 Terry Fox Drive, Ottawa,  ON  K2K 3J1, Canada} 

\author[0000-0002-8512-1404]{Takahiro Morishita}
\affil{Space Telescope Science Institute, 3700 San Martin Drive, Baltimore, MD 21218, USA}

\author[0000-0003-0409-0579]{Nathalie N.-Q. Ouellette}
\affiliation{Institut Trottier de recherche sur les exoplan\`etes, D\'epartement de physique, Universit\'e de Montr\'eal}
\affiliation{Observatoire du Mont-M\'egantic, Universit\'e de Montr\'eal, C.P. 6128, Succ. Centre-ville, Montr\'eal, H3C 3J7, Québec, Canada.}


\author[0000-0003-4196-0617]{Camilla Pacifici}
\affil{Space Telescope Science Institute, 3700 San Martin Drive, Baltimore, MD 21218, USA}

\author{Natasha Philippi}
\affiliation{Honeywell Aerospace \#100, 303 Terry Fox Drive, Ottawa,  ON  K2K 3J1, Canada} 

\author[0000-0002-3328-1203]{Michael Radica}
\affiliation{Institut Trottier de recherche sur les exoplan\`etes, D\'epartement de physique, Universit\'e de Montr\'eal}

\author[0000-0002-5269-6527]{Swara Ravindranath}
\affil{Space Telescope Science Institute, 3700 San Martin Drive, Baltimore, MD 21218, USA}

\author[0000-0002-5904-1865]{Jason Rowe}
\affiliation{Department of Physics \& Astronomy, Bishop's University, Sherbrooke, QC J1M 1Z7, Canada.}

\author[0000-0001-8127-5775]{Arpita Roy}
\affil{Space Telescope Science Institute, 3700 San Martin Drive, Baltimore, MD 21218, USA}
\affiliation{Department of Physics and Astronomy, Johns Hopkins University, 3701 San Martin Drive, Baltimore, MD 21218, USA}


\author{Karl Saad}
\affiliation{Canadian Space Agency, 6767 Route de l'Aéroport, Saint-Hubert, QC J3Y 8Y9, Canada}

\author[0000-0001-8368-0221]{Sangmo Tony Sohn}
\affil{Space Telescope Science Institute, 3700 San Martin Drive, Baltimore, MD 21218, USA}

\author[0000-0003-4787-2335]{Geert Jan Talens}\affil{Department of Astrophysical Sciences, Princeton University, 4 Ivy Lane, Princeton, NJ 08544, USA}


\author{Deepashri Thatte}
\affiliation{Space Telescope Science Institute, 3700 San Martin Drive, Baltimore, MD 21218, USA}

\author[0000-0003-4068-5545]{Joanna M. Taylor}
\affil{Space Telescope Science Institute, 3700 San Martin Drive, Baltimore, MD 21218, USA}

\author[0000-0002-5922-8267]{Thomas Vandal}
\affiliation{Institut Trottier de recherche sur les exoplan\`etes, D\'epartement de physique, Universit\'e de Montr\'eal}

\author[0000-0002-3824-8832]{Kevin Volk}
\affil{Space Telescope Science Institute, 3700 San Martin Drive, Baltimore, MD 21218, USA}

\author{Michel Wander}
\affiliation{Canadian Space Agency, 6767 Route de l'Aéroport, Saint-Hubert, QC J3Y 8Y9, Canada}

\author{Gerald Warner}
\affiliation{Honeywell Aerospace \#100, 303 Terry Fox Drive, Ottawa,  ON  K2K 3J1, Canada} 

\author{Sheng-Hai Zheng}
\affiliation{Honeywell Aerospace \#100, 303 Terry Fox Drive, Ottawa,  ON  K2K 3J1, Canada} 

\author{Julia Zhou}
\affiliation{Honeywell Aerospace \#100, 303 Terry Fox Drive, Ottawa,  ON  K2K 3J1, Canada} 

\author[0000-0002-4542-921X]{Roberto Abraham}
\affiliation{Department of Astronomy \& Astrophysics, University of Toronto, 50 St. George Street, Toronto, ON M5S 3H4, Canada}
\affiliation{Dunlap Institute for Astronomy and Astrophysics, University of Toronto, 50 St George Street, Toronto, ON M5S 3H4, Canada}

\author{Mathilde Beaulieu}
\affiliation{Université Côte d'Azur, Observatoire de la Côte d'Azur, CNRS, Laboratoire Lagrange, F-06108 Nice, France.}

\author{Bj\"orn Benneke} 
\affiliation{Institut Trottier de recherche sur les exoplan\`etes, D\'epartement de physique, Universit\'e de Montr\'eal}

\author[0000-0002-8224-1128]{Laura Ferrarese}
\affiliation{NRC Herzberg, 5071 West Saanich Rd, Victoria, BC V9E 2E7, Canada}

\author[0000-0001-5349-6853]{Ray Jayawardhana}
\affiliation{Department of Astronomy, Cornell University, Ithaca, New York 14853, USA}

\author[0000-0002-6773-459X]{Doug Johnstone}
\affiliation{NRC Herzberg, 5071 West Saanich Rd, Victoria, BC V9E 2E7, Canada}
\affiliation{Department of Physics and Astronomy, University of Victoria, 3800 Finnerty Road, Elliot Building, Victoria, BC, V8P 5C2, Canada}

\author[0000-0002-0436-1802]{Lisa Kaltenegger}
\affiliation{Carl Sagan Institute and Department of Astronomy, Cornell University, Ithaca, NY 14853, USA}

\author[0000-0003-1227-3084]{Michael R. Meyer}
\affil{Astronomy Department, University of Michigan, Ann Arbor, MI 48109, USA}

\author[0000-0002-0628-9605]{Judy L. Pipher}
\affiliation{Department of Physics and Astronomy, University of Rochester, Rochester NY 14627, USA}

\author[0000-0003-0029-0258]{Julien Rameau}
\affiliation{Institut Trottier de recherche sur les exoplan\`etes, D\'epartement de physique, Universit\'e de Montr\'eal}
\affiliation{Univ. Grenoble Alpes, CNRS, IPAG, F-38000 Grenoble, France.}

\author[0000-0002-7893-6170]{Marcia Rieke}
\affiliation{Steward Observatory, University of Arizona, 933 N. Cherry Ave, Tucson, AZ 85721, USA.}

\author[0000-0001-6758-7924]{Salma Salhi} 
\affiliation{Department of Physics and Astronomy, University of Calgary, 2500 University Dr NW, Calgary, AB T2N 1N4, Canada.}
\affiliation{Institut Trottier de recherche sur les exoplan\`etes, D\'epartement de physique, Universit\'e de Montr\'eal}

\author[0000-0002-7712-7857]{Marcin Sawicki}
\affil{Institute for Computational Astrophysics and Department of Astronomy \& Physics, Saint Mary's University, 923 Robie Street, Halifax, NS B3H 3C3, Canada}

\begin{abstract}

The Near-Infrared Imager and Slitless Spectrograph (NIRISS) is the science module of the Canadian-built Fine Guidance Sensor (FGS) onboard the James Webb Space Telescope (JWST). NIRISS has four observing modes: 1) broadband imaging featuring seven of the eight NIRCam broadband filters, 2) wide-field slitless spectroscopy (WFSS) at a resolving power of $\sim$150 between 0.8 and 2.2\,$\mu$m, 3) single-object cross-dispersed slitless spectroscopy (SOSS) enabling simultaneous wavelength coverage between 0.6 and 2.8\,$\mu$m at R$\sim$700, a mode optimized for exoplanet spectroscopy of relatively bright ($J<6.3$) stars and 4) aperture masking interferometry (AMI) between 2.8 and 4.8\,$\mu$m enabling high-contrast ($\sim10^{-3}-10^{-4}$) imaging at angular separations between 70 and 400 milliarcsec for relatively bright ($M<8$) sources. This paper presents an overview of the NIRISS instrument, its design, its scientific capabilities, and a summary of in-flight performance. NIRISS shows significantly better response shortward of $\sim2.5\,\mu$m resulting in 10-40\% sensitivity improvement for broadband and low-resolution spectroscopy compared to pre-flight predictions. Two time-series observations performed during instrument commissioning in the SOSS mode yield very stable spectro-photometry performance within $\sim$10\% of the expected noise. The first space-based companion detection of the tight binary star AB Dor AC through AMI was demonstrated.
\end{abstract}


\section{Introduction}\label{sec:intro}


One of the four science instruments onboard the James Webb Space Telescope (JWST; \cite{Gardner2006}, Mather et al. 2022) is the Near-Infrared Imager and Slitless spectrograph (NIRISS) provided by the Canadian Space Agency (CSA).
The original Canadian contribution to JWST was the Fine Guidance Sensor (FGS; \cite{Rowlands2003}) and technical contributions of the NIRCam instrument \citep{Horner2002}, more specifically the filter wheels and the tunable filter modules, used to be part of NIRCam in the early days of the Project and specifically designed to detect Lyman alpha emitters.  In 2002-2003, as part of the replan exercise of the Project that reduced the 8\,m telescope aperture into 6.5\,m, the tunable filter modules were displaced out of NIRCam to become a standalone instrument on the back side of FGS: the Tunable Filter Imager (TFI; \cite{Rowlands2004}).  However the original two channel TFI architecture was mass inefficient and was quickly re-optimized in 2005-2006 to a more compact single channel version which remained the baseline into Phase D \citep{Doyon2010}. The heart of TFI was a Fabry-Perot etalon whose development turned out to be challenging, in particular, their susceptibility to vibration and shocks during launch \citep{Haley2012}. Those technical issues led to a reconfiguration of TFI into a simpler, more robust instrument with imaging and spectroscopic capabilities enabling science programs complementary to the other three science instruments \citep{Doyon2012}.

This paper presents an overview of the NIRISS instrument, its observing modes, and a summary of in-flight performance based on commissioning data. This paper is part of a series describing the JWST mission \citep{Gardner2023}, the telescope performance \citep{McElwain2023}, the science performance \citep{Rigby2023} and that of the other three science instruments: NIRCam \citep{Rieke2023}, NIRSpec \citep{Boker2023} and MIRI \citep{Wright2023}. The reader is invited to consult the NIRISS section of the JDox documentation\footnote{\url{https://jwst-docs.stsci.edu}} for further information.

\section{Science requirements}\label{sec:instrument}
NIRISS was conceived to enable similar science themes as TFI which was designed for studying distant galaxies (Lyman $\alpha$ emitters) and for detecting/characterizing young gas giant exoplanets and brown dwarfs through narrow-band coronagraphy and aperture masking interferometry (AMI), the latter with a 7-hole non-redundant mask \citep{Doyon2010}. Since the reconfiguration had to occur over a relatively short schedule, approximately one year, this imposed minimal design changes that were restricted to keep the same optics, remove the Fabry-Pérot etalon and populate the dual (filter and pupil) wheel with new optical components (filters and dispersion elements). While the main intent of this reconfiguration was to provide new observing modes, they were also developed with the spirit of providing mode redundancy for broadband imaging to NIRCam, multi-object spectroscopy to NIRSpec and white light guiding to FGS.

\section{Observing Modes}

\begin{figure}[!htbp]
    \centering
    \includegraphics[width=\linewidth]{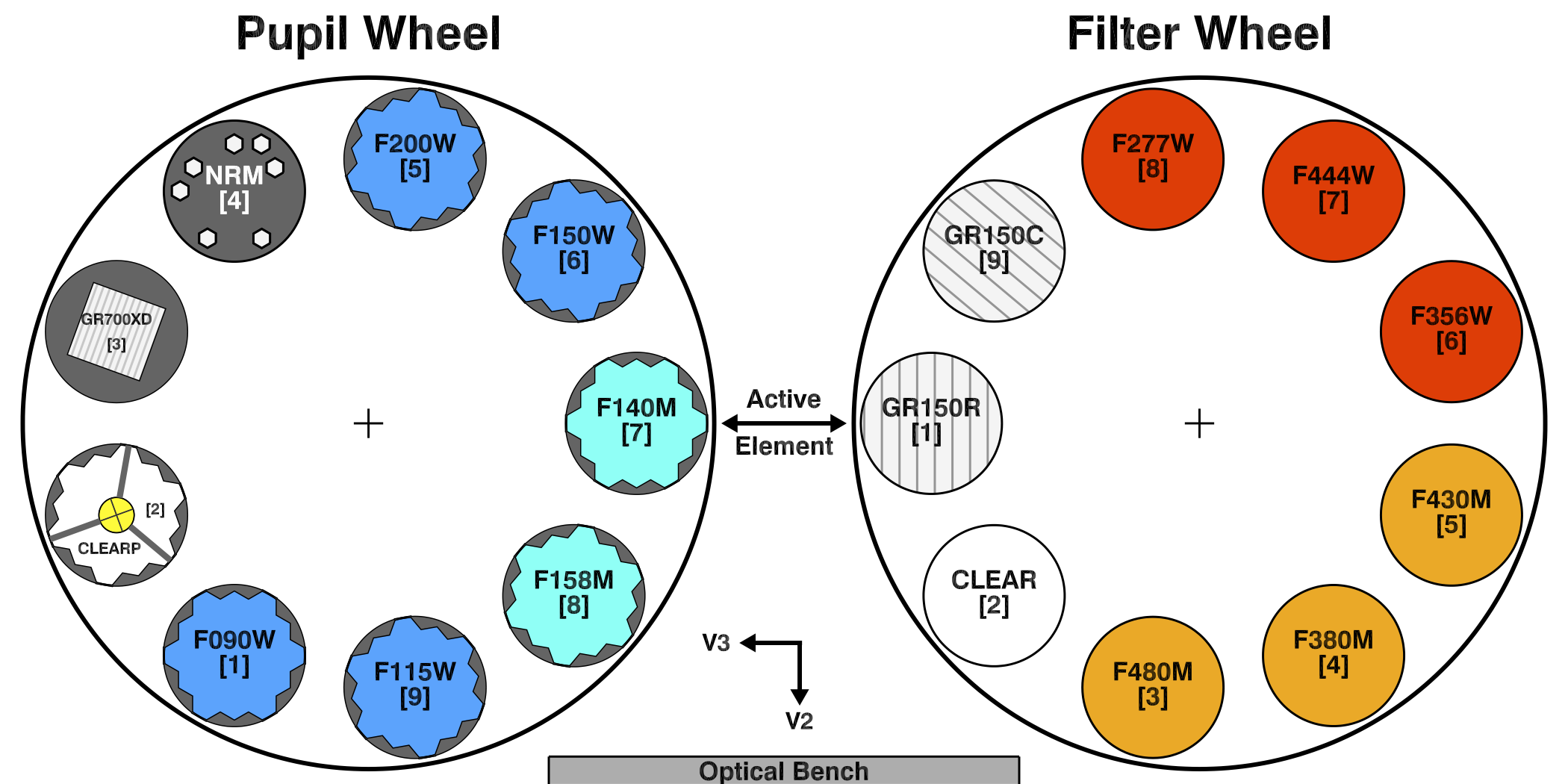}
    \caption{Schematic diagram showing the location of all optical elements on the filter and pupil wheels. Both wheels feature a clear aperture; CLEARP includes pupil alignment references in the central obstruction that was used during integration and tests. Figure from the JDox documentation.} 
    \label{fig:DualWheel}
\end{figure}

NIRISS features four observing modes: 1) broadband imaging, 2) low-resolution wide-field slitless spectroscopy (WFSS),  3) medium-resolution single-object spectroscopy (SOSS), and 4) aperture masking interferometry (AMI) and kernel phase interferometry (KPI). A given mode is configured by combining two optical elements from the filter and pupil wheels (see Figure~\ref{fig:DualWheel}). Here we present a brief description of each mode with highlights of the main science programs that they enable. More details on WFSS, SOSS, AMI and KPI are given in  \citet{Willott2022}, \citet{Albert2023}, \citet{Sivaramakrishnan2023} and \citet{Kammerer2023}, respectively. Table~\ref{table:ModeOverview} provides an overview of the NIRISS specifications along with a high-level description of each mode.

\subsection{Imaging}

\begin{figure}[!htbp]
    \centering
    \includegraphics[width=\linewidth]{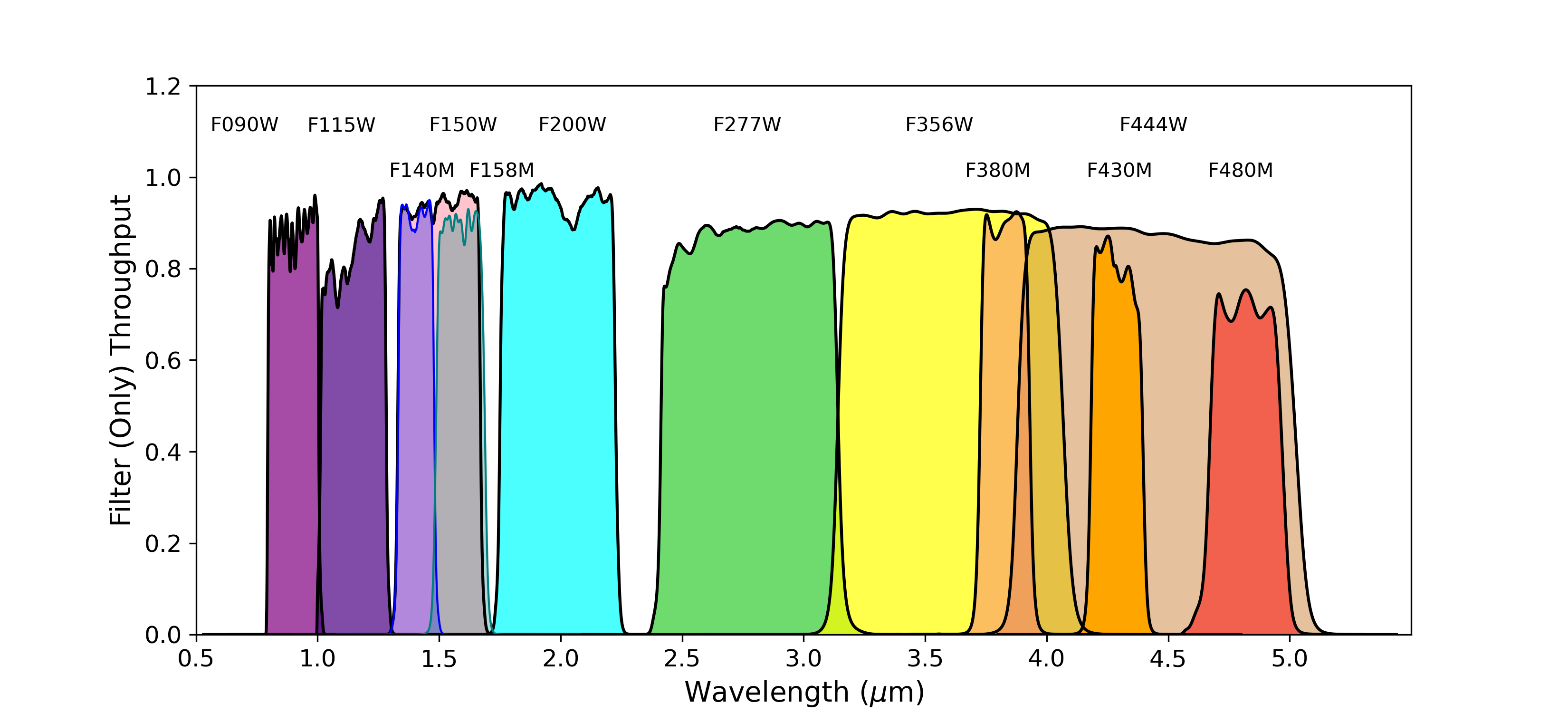}
    \caption{Transmission profile of all NIRISS filters. Figure from the JDox documentation.} 
    \label{fig:Filter_transmission}
\end{figure}

NIRISS imaging is primarily used for pre-imaging observations needed for the WFSS mode or for parallel observations but will be available as prime in Cycle 2. F070W excluded, NIRISS carries the same broadband ``W" filters as NIRCam with resolving power $R\sim4$. They were manufactured with the same coating design\footnote{Since the NIRCam short-wavelength channels use a 2.5\,$\mu$m cut-off H2RG detector, all filters shortward of 2.77\,$\mu$m are not blocked to radiation beyond 2.5\,$\mu$m.  For NIRISS, which uses a 5\,$\mu$m cut-off detector, each short-wavelength filter is mounted  with an appropriate blocking filter.} with inband transmission varying between 86 and 95\%  (see Figure~\ref{fig:Filter_transmission}). 

\begin{deluxetable}{lll}
\tablecaption{NIRISS Specifications and Observing Mode Overview }
\tablehead{
\colhead{Instrument Parameter} & \colhead{} & \colhead{Comment/Description}}
\startdata
Detector &  Hawaii-2RG & 5\,$\mu$m cut-off, 2048$\times$2048 pixels with 18\,$\mu$m pitch\\
Read noise (CDS) & 17.1  e$^-$& Total noise is 10 e$^-$ for a 1000\,s exposure\\
Image scale & 65.57$\pm$0.04 mas &  Distortion solution known to $<$3 mas RMS \\
Field of view & 2.2$^\prime\times 2.2^\prime$ & \\
\multicolumn{3}{c}{\textit{Imaging}}\\
Wavelength range & 0.8-5.0\,$\mu$m & 7 broad- and 5 medium-band filters \\
\multicolumn{3}{c}{\textit{WFSS}}\\
Wavelength range & 0.8-2.2\,$\mu$m & 6 blocking filters: F090W, F115W, F140M, F150W, F158M, F200W\\
Number of grisms & 2 & Two orthogonal orientations available \\
Spectral resolution & 139 & Measured 2-pixel resolution at 1.3\,$\mu$m\\
\multicolumn{3}{c}{\textit{SOSS}}\\
Wavelength range & 0.6-2.8\,$\mu$m &  Two cross-dispersed orders, no blocking filter \\
Spectral resolution & 654 & Measured 2-pixel resolution at 1.25\,$\mu$m\\
Brightness limit (J magnitude) & 6.3 & Order 2 in smallest subarray 96$\times$2048\\
\multicolumn{3}{c}{\textit{AMI}}\\
Wavelength range & 2.8-4.8\,$\mu$m & Four filters: F277W, F380M, F430M, F480M\\
Contrast$^a$ & 10$^{-3}$-10$^{-4}$ & One sigma contrast$\approx10/\sqrt{N_p}$ where $N_p$ is the total number of photons \\
Inner working angle & $70 - 400$\,mas & Total field of view is a subarray of  80$\times$80 pixels \\
\enddata
\tablecomments{$^a$ Preliminary in-flight performance are $0.5-1$\,mag worse than predicted with 10$^8$ photons \citep{Sivaramakrishnan2023}).}
\label{table:ModeOverview}
\end{deluxetable}

\subsection{Wide Field Slitless Spectroscopy}

The WFSS mode is optimized for the detection of faint, high-redshift galaxies over the full NIRISS field of view (FOV).  It is implemented through two grisms (GR150R \& GR150C) operated in slitless mode enabling low-resolution $R\sim150$  multi-object spectroscopy between 0.8 and 2.2\,$\mu$m in first order. 

Both grisms are identical except that they are mounted so as to disperse light 90$^\circ$ from one another, one along the detector rows (GR150R) and the other along the columns (GR150C). Observing the same scene with both grisms lifts the ambiguity between spatial location and wavelength, especially for emitting sources with strong equivalent-width such as Lyman $\alpha$ emitters or extreme emission-line galaxies. The two orientations also mitigate the problem of overlapping spectra for crowded regions for which the nominal operating scenario is to divide the observing time equally between GR150C and GR150R observations. The data are acquired similarly as in imaging mode by dithering the scene around on the detector.  

\subsection{Single-Object Slitless Spectroscopy}
SOSS is specifically designed to perform medium-resolution ($R\sim700$) transit and eclipse spectroscopy of exoplanets orbiting relatively bright ($J>6.3$) host stars. 
The wide wavelength coverage is achieved through the GR700XD grism assembly featuring a cross-dispersing ZnS prism combined with a directly-ruled ZnSe grism (c.f., figure 1 of \citet{Albert2023}). The entrance face of the ZnS prism has a built-in cylindrical weak lens that defocuses the spectrum over $\sim$25 pixels along the spatial direction without degrading the spectral resolution, effectively increasing the brightness limit by $\sim$3 magnitudes. Because of mechanical constraints, i.e. clearance between the dual wheel and the optomechanical bench, the GR700XD could not be made thick enough to separate all orders completely. As a result, order 2 is slightly overlapping with order 1 around 2.2\,$\mu$m. The SOSS data pipeline features special algorithms for locating all spectral traces \citep{Radica2022} and to correct the inter-order contamination that turns out to be of minimal scientific impact even uncorrected \citep{Darveau-Bernier2022}.

\subsection{Aperture Masking Interferometry}

AMI enables high-contrast (10$^{-3}-10^{-4}$) imaging between 2.8 and 4.8\,$\mu$m at inner working angles (IWA) between 70 and 400 mas. Such high-contrast sub- $\lambda/D$ imaging capability (\citet{Sivaramakrishnan2022} and references therein) is unique and very complementary to the NIRCam coronagraph that includes occulting spots with IWA between 300 and 880 mas over the same wavelength range \citep{Rieke2023}. AMI probes a parameter space (IWA$<$200 mas  at 4-5\, $\mu$m) hardly accessible from ground-based 8-10\,m class instruments like the Gemini Planet Imager \citep{Macintosh2014}, SPHERE \citep{Beuzit2019} and NIRC2. 

AMI is implemented through a 7-hole non-redundant mask (see Figure~\ref{fig:DualWheel}) optimized for detecting point sources, but it can also be used for extended sources such as transition disks and active galactic nuclei; examples of such applications are given in \cite{Sivaramakrishnan2023}. AMI can be used with F277W and three medium-band ($R=16-20$) filters (F380M, F430M, F480M) whose wavelengths were specifically chosen to provide good constraints on both  the effective temperature and surface gravity of relatively cold exoplanets and brown dwarfs with clear atmospheres (see Figure 13 from \citet{Doyon2012}). 

KPI (\citet{Martinache2010}) is the full aperture version of AMI, i.e. without an  NRM mask that also provides sub-$\lambda/D$ imaging but for fainter targets albeit with lower contrasts. This mode is particularly appealing for detecting close binary systems like brown dwarfs. This mode is not specific to NIRISS and can be applied to NIRCam and MIRI but only NIRISS was used for testing KPI during commissioning. Detailed results from NIRISS KPI observations are presented in \citet{Kammerer2023}. 


\section{Instrument Design}
NIRISS was designed, manufactured, assembled and tested by Honeywell Aerospace (Ottawa ON, Canada) under contract to CSA, with contributions from CSA, the Université de Montréal, and the Herzberg Astronomy \& Astrophysics Research Centre of the National Research Council Canada (Victoria, BC).

\subsection{Optical Design}\label{sec:optical design}
\begin{figure}[!htbp]
    \centering
    \includegraphics[width=\linewidth]{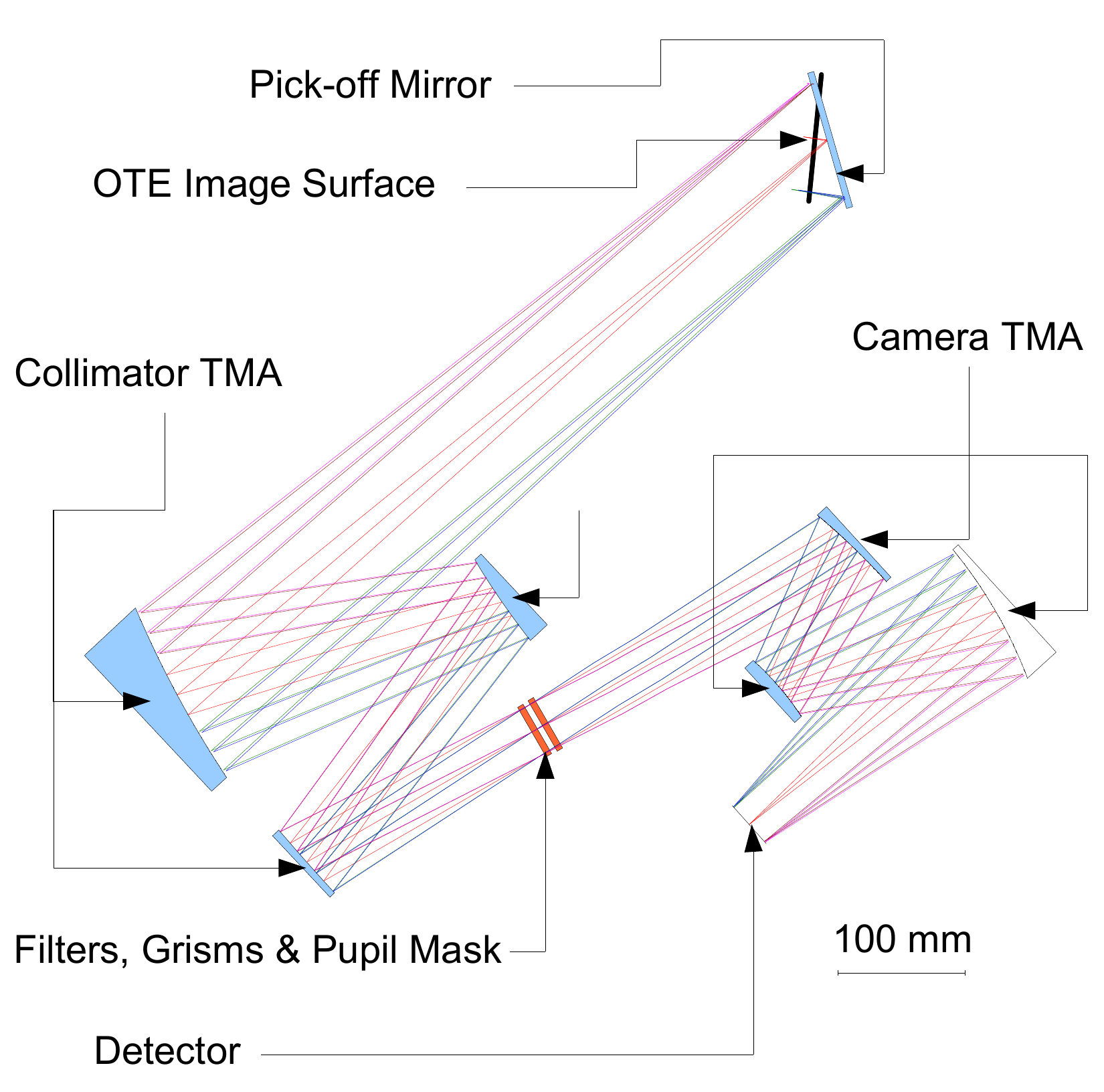}
    \caption{Schematic of the NIRISS optical design.}
    \label{fig:Optical-layout}
\end{figure}

NIRISS features an all-reflective optical design based on aspheric diamond-turned Aluminium mirrors manufactured by Corning NetOptix (Keene NH, USA). A layout of the optical design is shown in Figure \ref{fig:Optical-layout}. The NIRISS FOV is captured by a pick-off (POM) mirror located at the telescope focus; the POM is mounted on a linear stage to allow for potential focus adjustment of up to $\pm$5.2 mm in the OTE focal plane.  After the highly successful OTE commissioning campaign the telescope focal position was found to provide excellent image quality and no focus adjustment of the NIRISS POM was needed \citep{McElwain2023}. The light then passes through a three-mirror assembly (TMA) whose output is a 39-mm diameter collimated beam that feeds the dual wheel (pupil then filter wheel) followed by a TMA camera that refocuses light onto the detector at F/8.8 to yield the nominal 65\,mas image scale. The wavefront error (WFE) of the whole optical train was measured at cryogenic temperature and varies between 50 and 70\,nm depending on the FOV point.

\subsection{Mechanical Design}\label{sec:mechanical design}
Both FGS and NIRISS share a common optomechanical bench made of Aluminium to ease optical alignment from room to cryogenic temperature. Three titanium kinematic mounts provide the interface to the carbon fibre structure of JWST’s Integrated Science Instrument Module (ISIM). A mechanical layout of the NIRISS side (without baffles) is shown in Figure \ref{fig:NIRISS-CAD} along with a flight hardware photograph. Figure \ref{fig:NIRISS-photos} shows  photographs of some of the instrument's key subsystems: the detector focal plane assembly (FPA) and the dual wheel. 

\subsection{Electrical/Software Design}\label{sec:electrical design}
\begin{figure*}[!htbp]
    \centering
    \includegraphics[width=\linewidth]{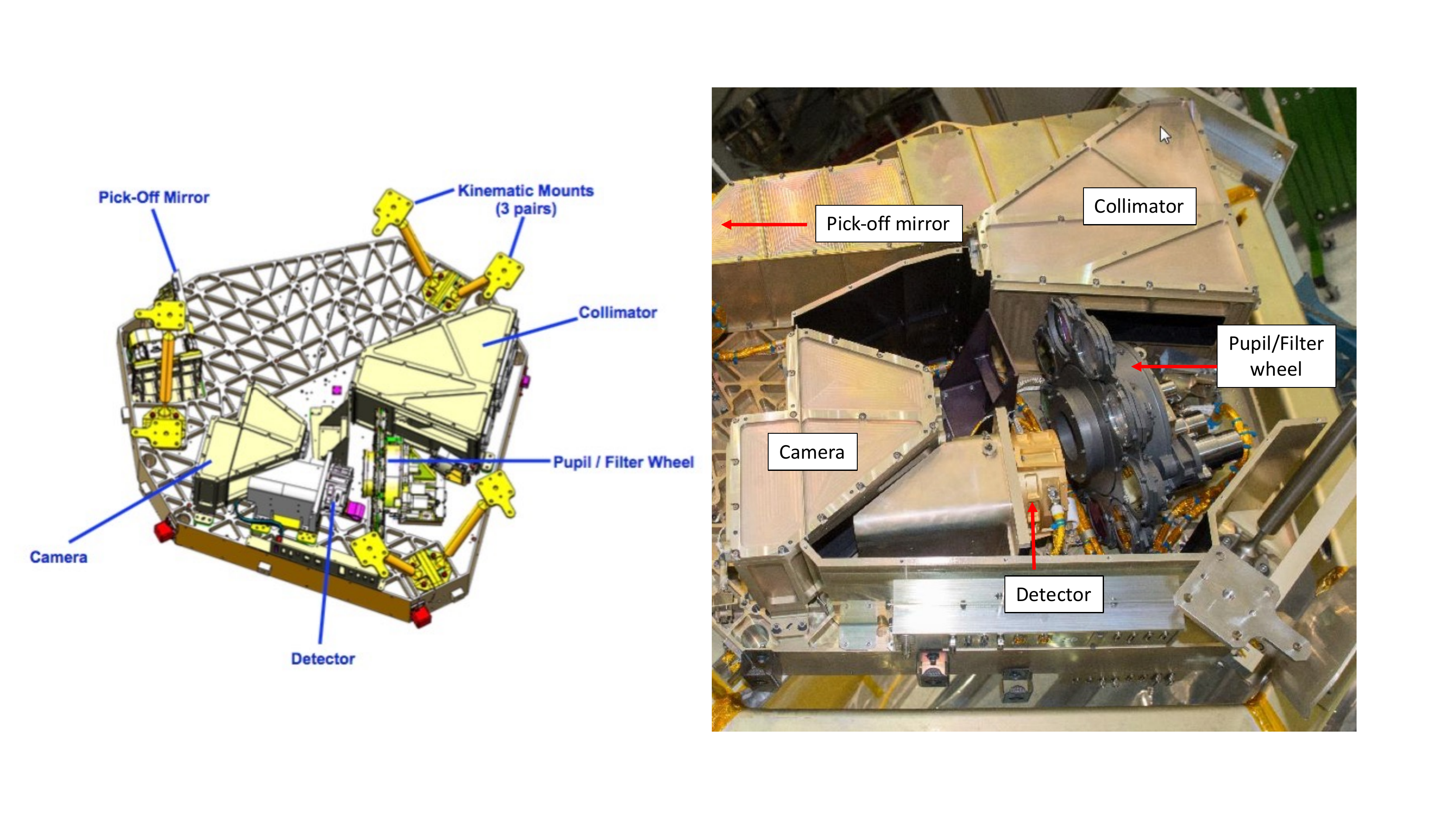}
    \caption{Left: solid rendering of the NIRISS optomechanical bench shown without the baffle between the pick-off mirror and the collimator TMA. Right: flight hardware photograph. }
    \label{fig:NIRISS-CAD}
\end{figure*}

\begin{figure*}[!htbp]
    \centering
    \includegraphics[width=\linewidth]{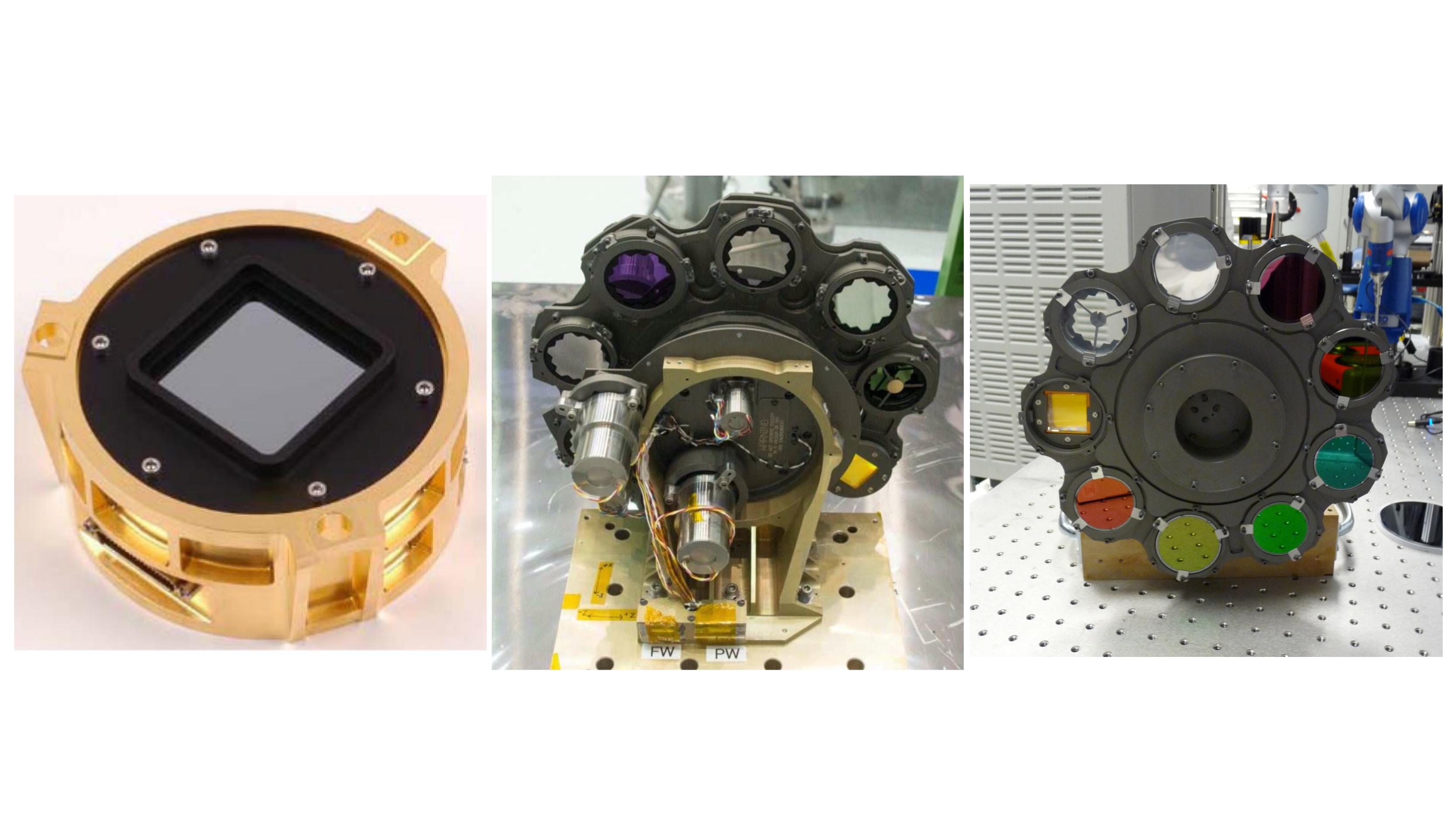}
    \caption{Left: focal plane assembly unit, center: pupil side of the dual wheel and the filter wheel on the right.}
    \label{fig:NIRISS-photos}
\end{figure*}

In addition to the FGS/NIRISS optical assembly (Figure \ref{fig:NIRISS-CAD}), the other major hardware component of NIRISS is the FGS/NIRISS electronics box located in the ISIM Electrical Compartment (IEC).  The IEC is located beside the cryogenic ISIM structure supporting all the Science Instruments and is maintained at a temperature of $\sim$280 K.  Special radiators ensure that the thermal load on the $\sim$ 40 K ISIM structure is minimized. The FGS/NIRISS Electronics Box is organized as three independent sets of control electronics, one for each guider and one for the (single string) NIRISS instrument.  Each guider and NIRISS has four identical printed circuit assemblies (a power control card, a Spacewire interface card, a SIDECAR control card and a Focus Mechanism \& Housekeeping card.  In addition, the NIRISS electronics box has a Dual Wheel control card and a ‘blank’ card to terminate the control signals originally intended for the Fabry-Perot etalon in the former TFI configuration.      

Instrument control functions are performed by NIRISS Flight Software (FSW) which is hosted in the ISIM Command \& Data Handling Unit.  Communication with the FGS/NIRISS electronics is solely over a redundant Spacewire interface, unlike the other JWST science instruments that make use of a 1553 interface for command \& telemetry and a Spacewire interface for image data flow.    All FGS/NIRISS mechanisms make use of stepper motors with resolvers to read back their current positions.  The gearing of the filter and pupil wheels is such that the resolver completes one rotation as the wheel moves through each filter/pupil position.  The wheels also have a variable reluctance (VR) sensor which triggers at the home positions of the wheels, in order for NIRISS FSW to confirm filter/pupil position. 

NIRISS FSW also provides temperature control for the focal plane array and, if required, the SIDECAR ASIC.   The FPA is maintained at $\sim$39 K, with $\sim$20 mK of drift depending on the thermal state of the observatory.  The FPA temperature can also drop by $\sim$30 mK during sub-array readouts.

NIRISS FSW also controls four redundant miniature incandescent lamps which project into the NIRISS optical path upstream of the filter and pupil optical elements.  The illumination of the focal plane is non-uniform and the resulting calibration image data is used to monitor relative changes.   Two of the lamps project through a comb filter (intended for the TFI configuration) and two are unfiltered.

\section{In-flight NIRISS Performance}
This section provides a high-level summary of the instrument performance solely based on commissioning data that are compared to pre-flight measurements. We will show that NIRISS performs significantly better than predictions in general.

\subsection{Detector}

The NIRISS detector is controlled by an Application-Specific Integrated Circuit (ASIC)  called SIDECAR \citep{Loose2006} which provides biases, clocking and output digitization.  The SIDECAR bias voltages affect the read noise, dark current and dynamic range of the detector, thus best performance is obtained when the detector and SIDECAR are optimized together.  The flight detector for NIRISS was characterized at Teledyne in their standard test set, then more fully characterized at Honeywell with a representative SIDECAR ASIC. The final optimization was done
through an ASIC tuning procedure during the CV3 test campaign in 2015. This procedure was not repeated on orbit since the performance was very similar to during ground tests. The last pre-flight detector tests were obtained during the OTIS test campaign in 2017 when the whole observatory was tested at cryogenic temperature. We take the OTIS data as our reference for pre-flight detector performance.  

\begin{figure}[!htbp]
    \centering
    \includegraphics[width=\linewidth]{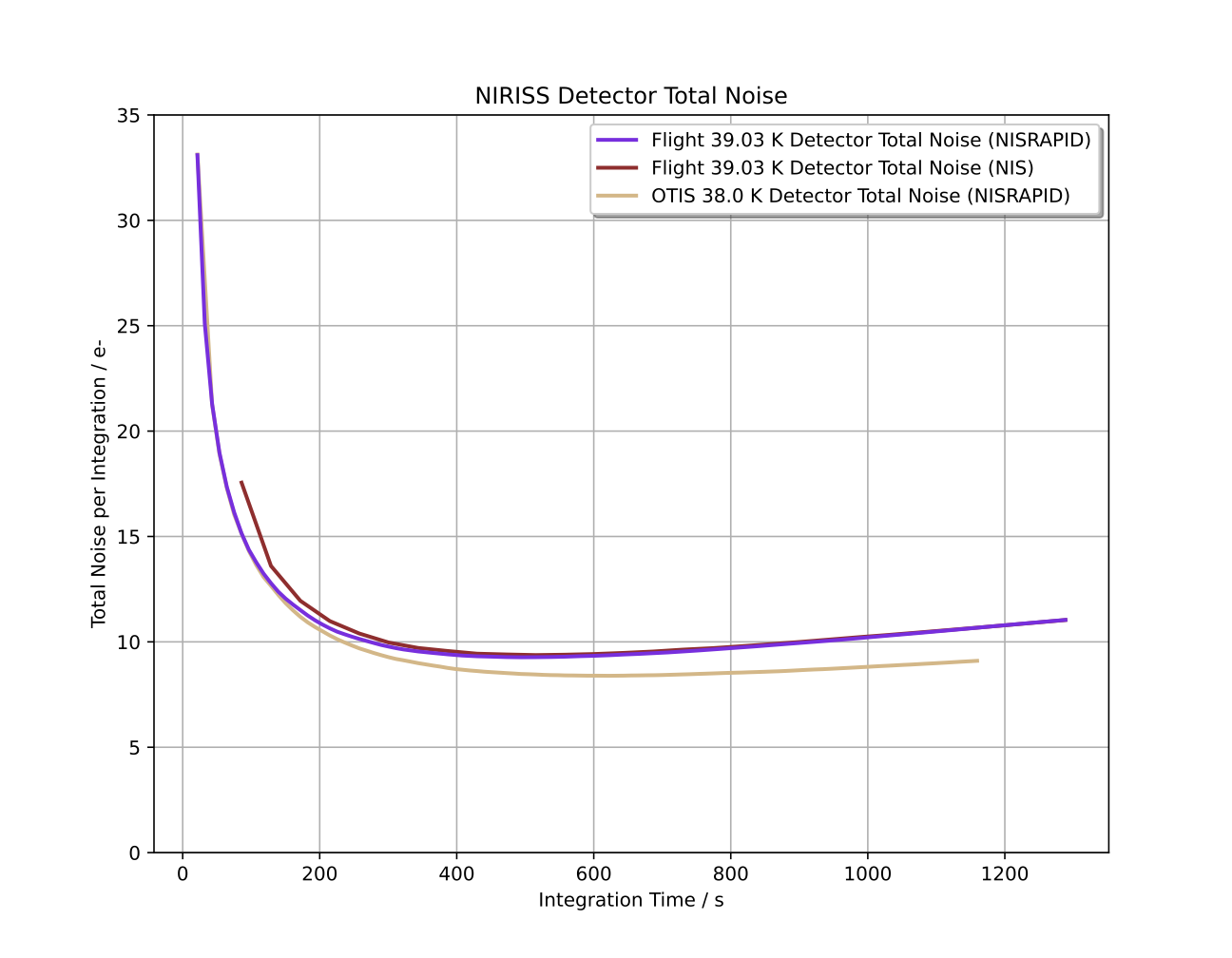}
    \caption{Total noise vs exposure time. The on-orbit performance is slightly worse than pre-flight. The most likely cause is uncorrected cosmic rays.} 
    \label{fig:Total-noise}
\end{figure}

The NIRISS detector has a distinctive feature associated with an epoxy void. While most of the Teledyne H2RG arrays on JWST have voids, the NIRISS void is the largest, occupying ~8\% of the detector area (see Figure 2 of \citet{Albert2023}). This void region has a dark current about a factor of two lower than the rest of the detector. The detector performance is summarized in Table \ref{table:Detector-performance}. The in-flight correlated double sampling (CDS) read noise is marginally lower compared to that at OTIS but the dark current is significantly higher by 0.005 e-/s. This increase is most likely associated with cosmic ray residuals not properly handled by the jump detection algorithm during the up-the-ramp sampling. This higher dark current probably explains the slightly higher in-flight total noise (see Figure \ref{fig:Total-noise}). This higher dark current has a marginal scientific impact since relatively long integrations from WFSS programs are background-limited.

\begin{deluxetable}{lll}
\tablecaption{In-flight and Pre-flight (OTIS) Detector Performance$^{\rm a}$}
\tablehead{
\colhead{Median stats} & \colhead{In-flight (39.0 K) } & \colhead{OTIS (38.0  K)}}
\startdata
Dark current outside void e$^-$/s$^{\rm b}$ & 0.048 & 0.043 \\
Dark current inside void e$^-$/s$^{\rm b}$ & 0.026 & 0.021 \\
CDS Noise e-/s$^{\rm b}$ & 17.06 & 17.25 \\
Amplifier Crosstalk & 0.00028 & 0.00021 \\
Inter-Pixel Capacitance (IPC) & 0.00464 & 0.00432 \\
Number of hot ($>$0.5 e$^-$/s) pixels & 6257 &  5874 \\
Number of warm (0.1-0.5 e$^-$/s) pixels & 8494 & 7071 \\
Number of noisy ($>$30\,ADU) pixels & 3072 & 3306 \\
\enddata
\tablecomments{$^{\rm a}$ Full-frame measurements with the NISRAPID read out mode. \\
$^{\rm b}$ Assumes the measured conversion gain of 1.62 e$^-$/ADU (analog to digital unit).}
\label{table:Detector-performance}
\end{deluxetable}

\subsection{Image Quality}

\begin{figure}[!htbp]
    \centering
    \includegraphics[width=\linewidth]{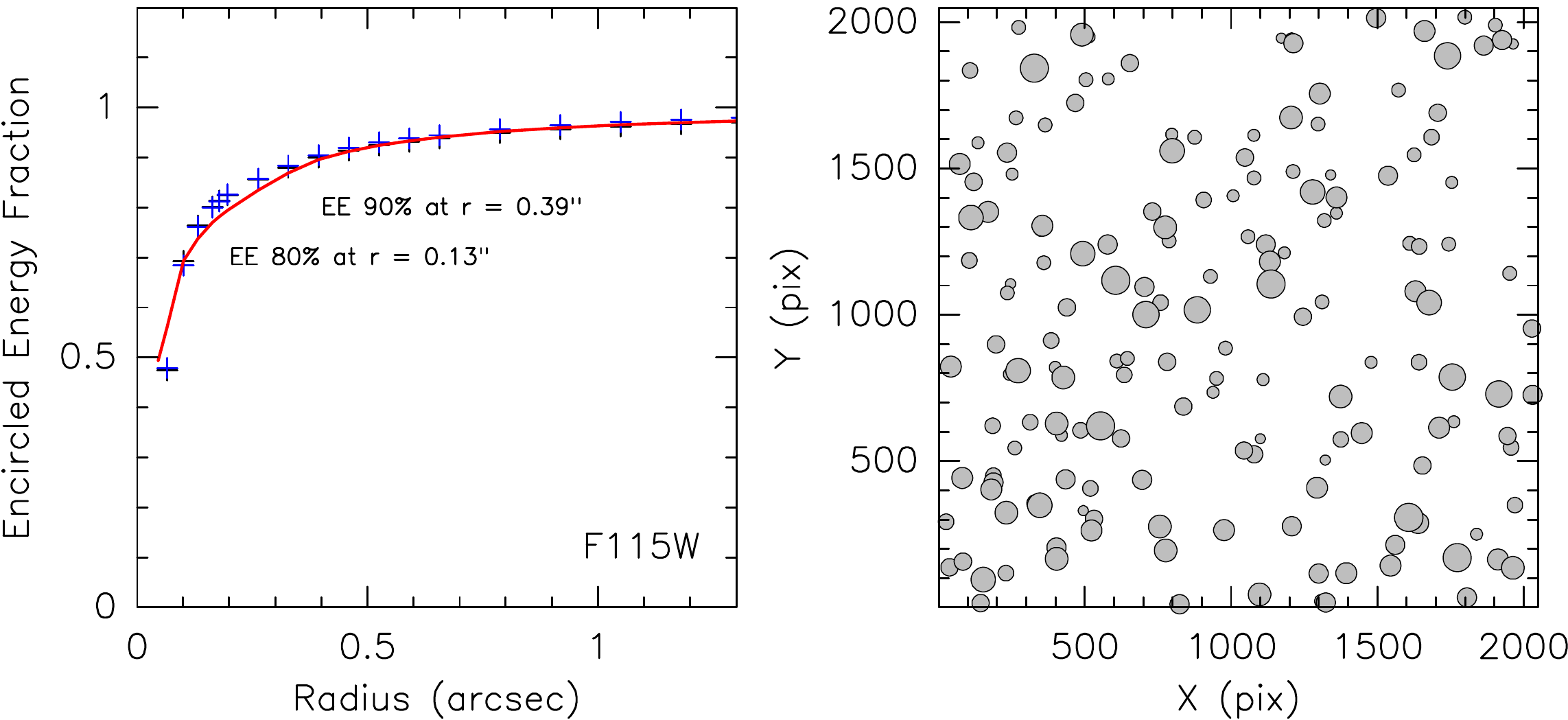}
    \caption{{\it Left panel}: encircled energy (EE) as a function of measurement radius for filter F115W. Plus signs in black and blue represent measurements for two different dither positions. For comparison, the red curve represents the run of EE for the pre-launch WebbPSF model for F115W. The radii for EE levels of 80\% and 90\% are indicated. {\it Right panel}: ellipticity measurements of stars across the field of view for exposures with the F480M filter in commissioning program 1464. The symbol size indicates the ellipticity, ranging from 0.01 to 0.11. The mean ellipticity in this image is 0.03 $\pm$ 0.01. Note the lack of any trend in ellipticity across the FOV.}
  
    \label{fig:ImageQuality2}
\end{figure}

\begin{figure}[!htbp]
    \centering
    \includegraphics[width=\linewidth]{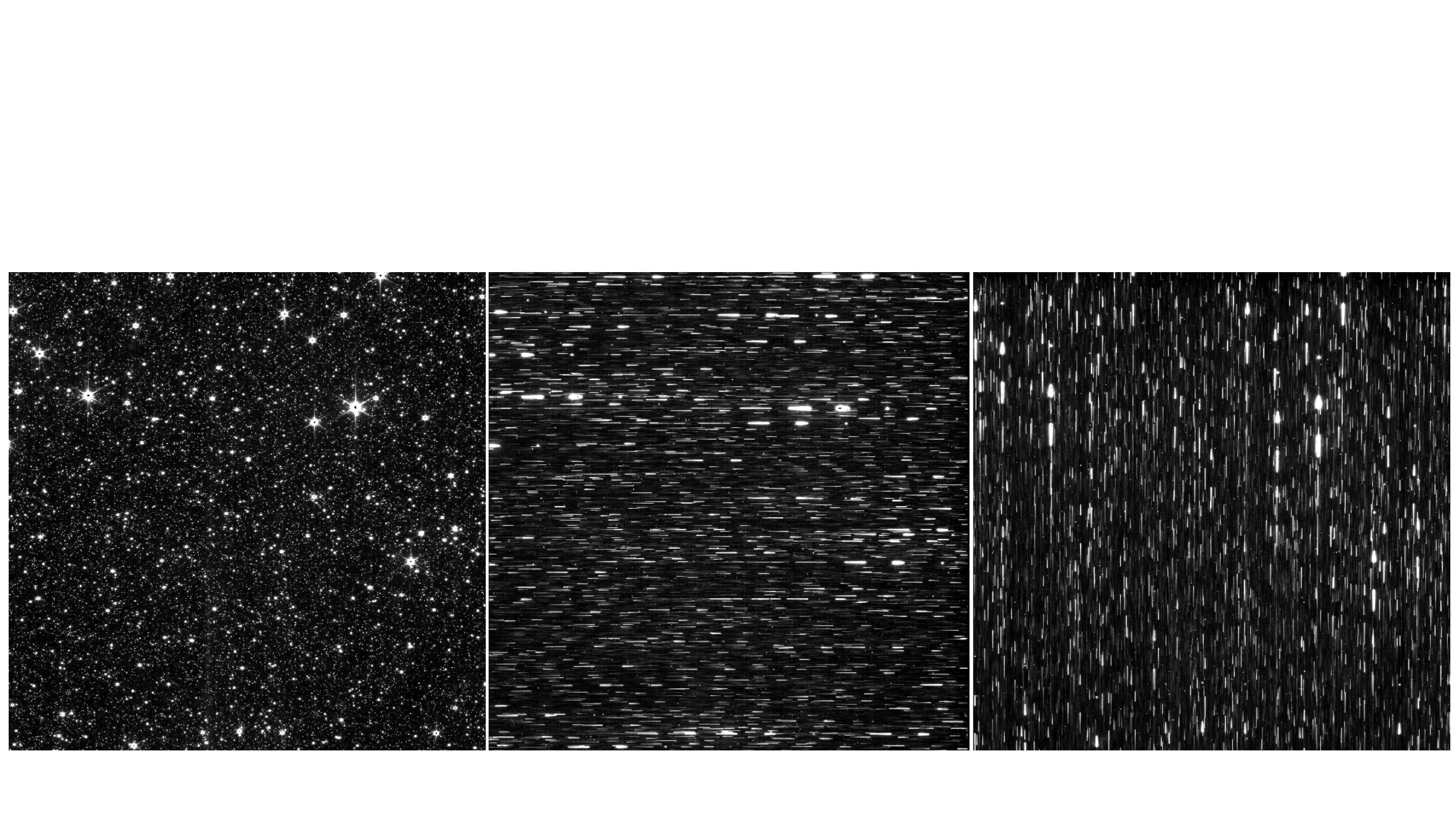}
    \caption{{\it Left}: F200W image of a crowded star field, illustrative of the very uniform PSF across the whole FOV. {\it Middle and right}: Same field with the GR150C/R grisms.  } 
    \label{fig:WFSS-LMC}
\end{figure}

NIRISS image quality was characterized by measuring the encircled energy (EE), ellipticity and full width at half maximum (FWHM) of the point spread function (PSF) of bright stars at several FOV locations. As shown in Figure \ref{fig:ImageQuality2}, the measured EE exceeds pre-flight  predictions from WebbPSF with hardly measurable PSF variations across the FOV. Figure~\ref{fig:WFSS-LMC} is a qualitative illustration of the excellent image quality for both imaging and WFSS. Such an exquisite image quality performance is due to the excellent telescope wavefront error measured to be diffraction-limited at 1.1\,$\mu$m instead of the 2\,$\mu$m requirement \citep{Menzel2023}.  

\subsection{Throughput}
The system throughput was measured for all filters, the GR150C/R and GR700XD grisms. As presented below, all modes show a throughput improvement between 10 and 60\% compared to the pre-flight Exposure Time Calculator (ETC).
This improved transmission is likely attributed to several factors: a combination of better throughput from the telescope optics, the NIRISS TMA, the quantum efficiency of the detector and the accumulation of margins in the throughput budget.  

\subsubsection{Imaging}
Table \ref{table:ImagingThroughput} reports the measured count rates relative to the pre-flight ETC and the instrument team predictions. Overall, the measured throughput is 10-20\% better than predicted. Both the pre-flight ETC and the instrument team predictions are in good agreement in general except for filters longward of $\sim2\,\mu$m, a discrepancy that has now been routed to an aperture correction issue within the pre-flight ETC that has been fixed for the version available from December 2022 onwards.

\begin{deluxetable}{lll}
\tablecaption{Measured count rates relative to predictions$^{\rm a}$}
\tablehead{
\colhead{Filter} & \colhead{$\tau_{ETC}(\lambda)^{\rm b}$ } & \colhead{$\tau_{NIS}(\lambda)^{\rm c}$}}
\startdata
F090W & 1.25 & 1.21 \\
F115W & 1.26 & 1.21 \\
F140M & 1.31 & 1.25 \\
F150W & 1.29 & 1.23 \\
F158M & 1.29 & 1.2 \\
F200W & 1.14 & 1.07 \\
F277W & 1.22 & 1.11 \\
F356W & 1.24 & 1.10 \\
F380M & 1.25 & 1.07 \\
F430M & 1.23 & 1.07 \\
F444M & 1.20 & 1.13 \\
F480M & 1.27 & 1.08 
\enddata

\tablecomments{$^{\rm a}$ Measured from the flux standard LDS749B.\\
$^{\rm b}$ Measured count rate relative to pre-flight ETC predictions.\\
$^{\rm c}$ Measured count rate relative to NIRISS instrument team pre-flight predictions.}
\label{table:ImagingThroughput}
\end{deluxetable}

\subsubsection{WFSS}
\begin{figure}[!htbp]
    \centering
    \includegraphics[width=\linewidth]{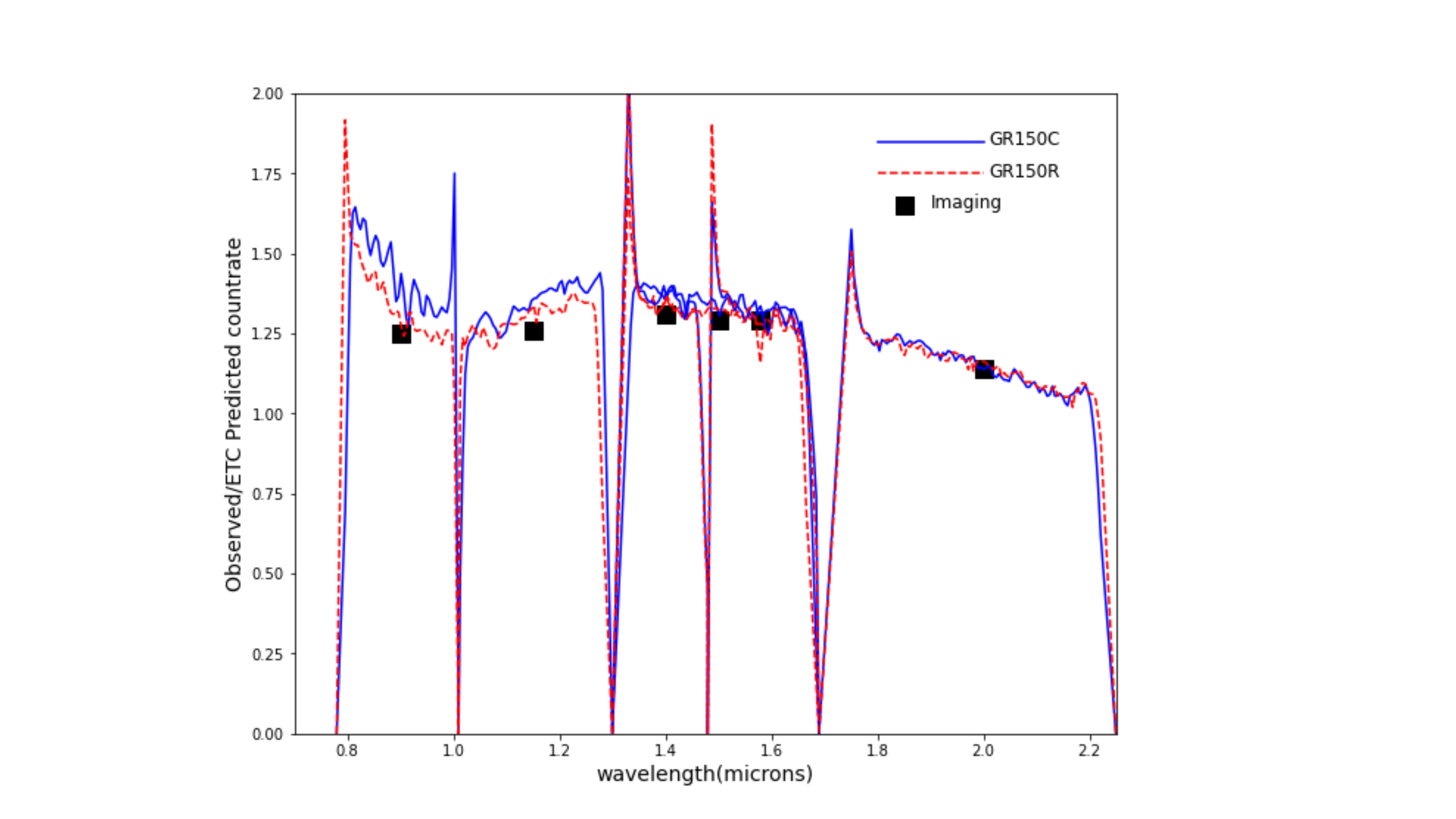}
    \caption{Measured count rates of both GR150 grisms relative to pre-flight ETC predictions. Black squares are measurements inferred from imaging data.} 
    \label{fig:WFSS-Throughput}
\end{figure}

The throughput of both GR150 grisms was characterized over all short wavelength filters by observing the spectroscopic standard WD\,1657+343. As shown in Figure \ref{fig:WFSS-Throughput}, the throughput of both grisms is significantly better than the pre-flight ETC predictions by 10-25\%, as high as 50\% near $0.8\,\mu$m, in excellent agreement with the imaging data. GR150C has a slightly better transmission at $<1.3\,\mu$m than GR150R, most likely due to a better coating. GR150C should thus be preferred for observations requiring only one orientation. 

The absolute transmission of both grisms was measured by comparing the relative signal with/without blocking filters on low background fields (see section~\ref{section:background}). The resulting transmission is $85\pm3$\%, in good agreement with pre-flight measurements. 

\subsubsection{SOSS}

\begin{figure*}[!htbp]
    \centering
    \includegraphics[width=\linewidth]{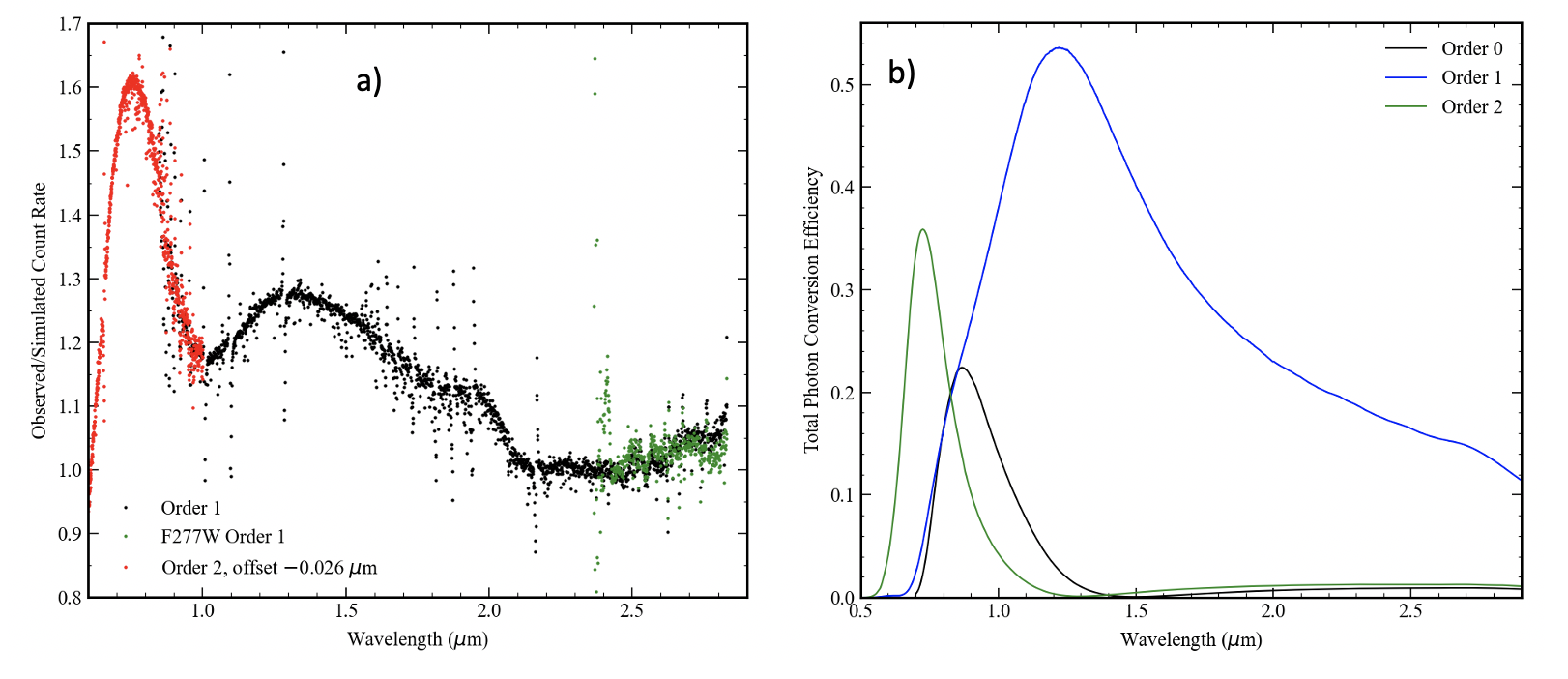}
    \caption{a) Measured divided by the predicted count rate of the SOSS grism in both order 1 (black) and 2 (red), the latter shifted by 0.026 $\mu$m to match the pre-flight blaze function with in-flight data. b) Resulting photon conversion efficiency.} 
    \label{fig:SOSS-Throughput}
\end{figure*}

The GR700XD grism was characterized by observing the A1V star BD+60 1753 ($J=9.61$) with and without the F277W filter, the latter allowing to eliminate order 2 in the overlap region. As shown in Figure \ref{fig:SOSS-Throughput}a, order 1 shows a throughput ~25\% better than the pre-flight ETC predictions around the blaze peak (1.3 $\mu$m).  Compared to the ETC, order 2 shows an improved transmission of $\sim$60\% now understood to be a slight mismatch of 0.026 $\mu$m between the in-flight and pre-flight measurements of the blaze wavelength. Figure~\ref{fig:SOSS-Throughput}b shows the resulting photon conversion efficiency of the GR700XD for both orders. Order 2 shows a significant transmission of $\sim$24\% near H$\alpha$ (0.656 $\mu$m) which is a very good stellar activity indicator for low-mass stars.

\subsection{Wavelength Calibration}

\subsubsection{WFSS}

\begin{figure}[!htbp]
    \centering
    \includegraphics[width=\linewidth]{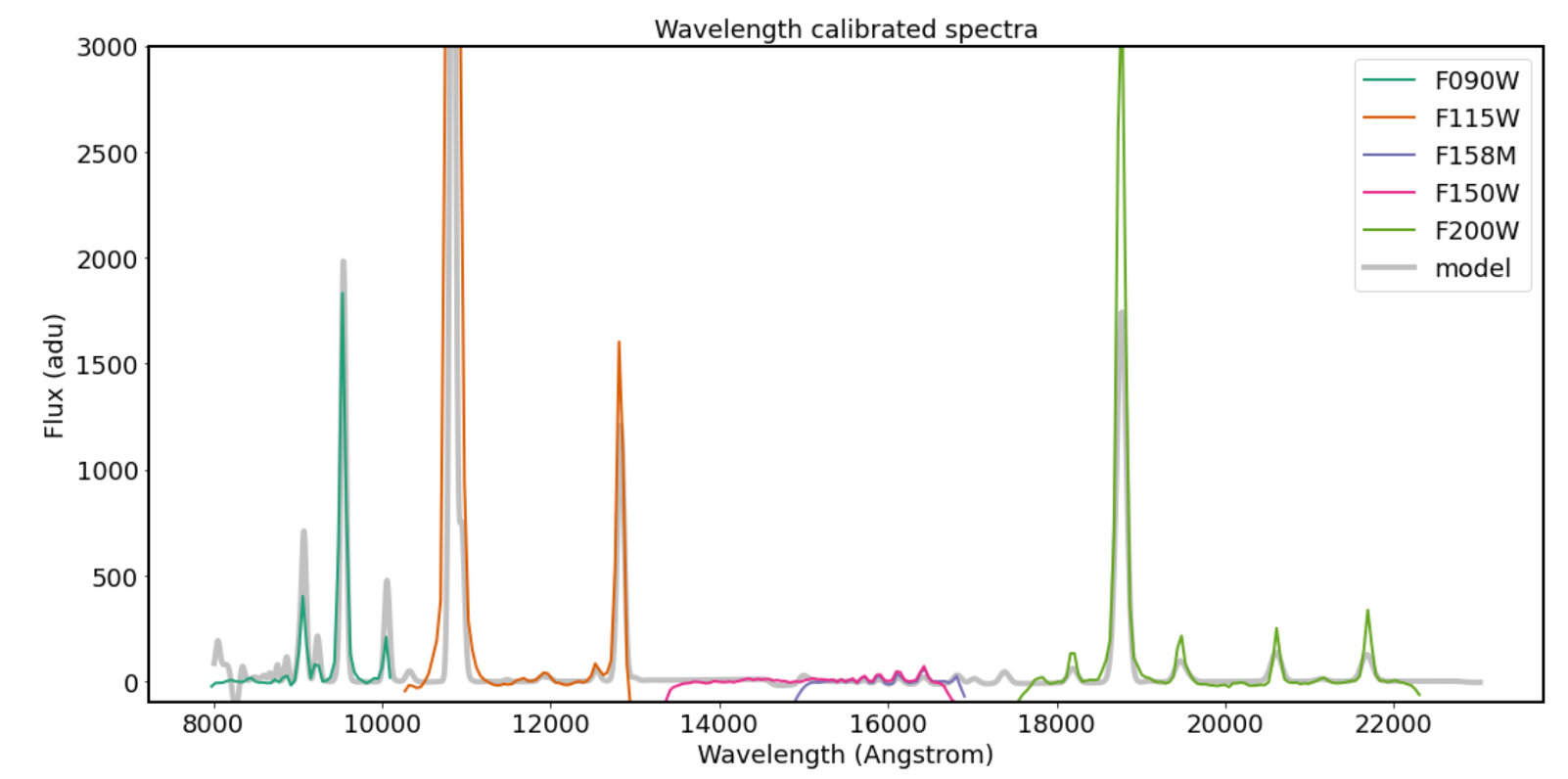}
    \caption{GR150R spectrum of a planetary nebula observed with all short-wavelength filters.} 
    \label{fig:WFSS-wavecal}
\end{figure}

The wavelength calibration of both GR150 grisms was achieved by observing the compact planetary nebula SMP-LMC-58.  Figure~\ref{fig:WFSS-wavecal} presents a representative spectrum with the GR150R grism, showing an excellent agreement with the model. The inferred linear dispersions are $46.888\pm 0.009\,\AA/{\rm pix}$ and $46.959\pm 0.017\,\AA/{\rm pix}$ for the GR150R and GR150C grims, respectively, in excellent agreement with the optical model predictions. This dispersion corresponds to a 2-pixel resolving power of 139 at 1.3 $\mu$m.

\subsubsection{SOSS}

\begin{figure}[!htbp]
    \centering
    \includegraphics[width=\linewidth]{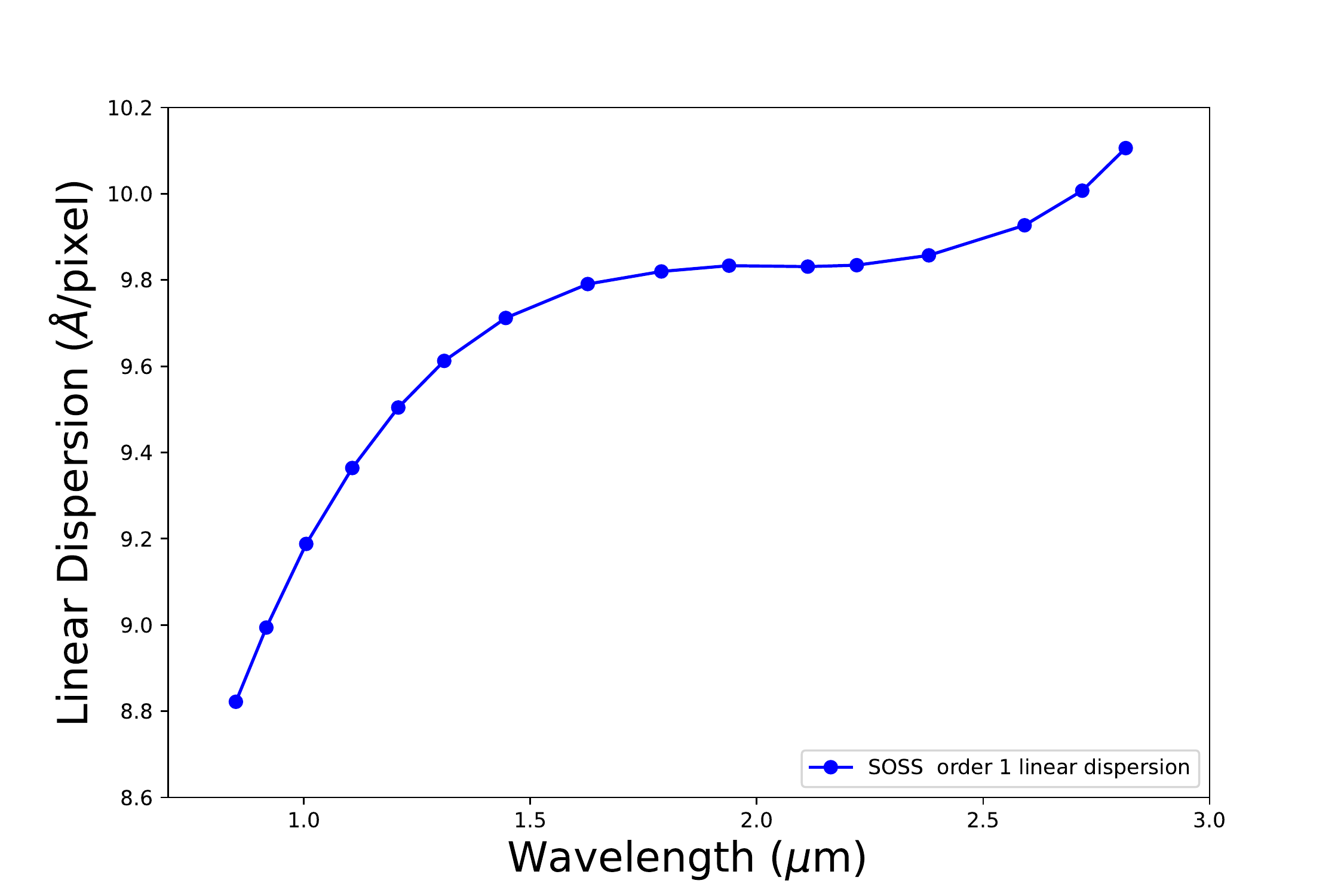}
    \caption{Linear dispersion of the SOSS order 1.} 
    \label{fig:SOSS-dispersion}
\end{figure}

    The SOSS wavelength calibration was achieved through spectral correlation with the following stars: BD+601753 (A1V) and TWA33 (M5.5V). Hydrogen lines were used to anchor the wavelength solution from the A star while a BT-Settl atmosphere model \citep{Allard2012} was used to derive the one from the M dwarf. The dispersion shows some non-linearity (see~Figure \ref{fig:SOSS-dispersion}) due to optical distortion within the SOSS FOV. The inferred median dispersions are $9.8\pm0.3$\,\AA/pixel and $4.6\pm0.2$\,\AA/pixel for orders 1 and 2, respectively, with corresponding median 2-pixel resolving powers of $\sim$700 at blaze peak (1.3 $\mu$m for order 1 and 0.7\,$\mu$m for order 2). The current wavelength solution is accurate to $\sim$1/3 pixel over most of the SOSS spectral range with some systematic mismatch at the level of $\sim$2 pixels around 2.3 $\mu$m. The wavelength solution should improve with more data during Cycle 1.

\subsection{Astrometric calibration}

The plate scale and optical distortion of the NIRISS FOV were characterized through observations within an astrometric field located in the Large Magellanic Cloud, a field extensively used and characterized for the Hubble Space Telescope (see the \href{https://jwst-docs.stsci.edu/jwst-data-calibration-considerations/jwst-data-absolute-astrometric-calibration}{Astrometric Calibration} section of the JDox documentation.). The resulting astrometric solution is accurate to 3 milliarcsec and repeatable to better than 1 milliarcsec. The measured plate scale is $65.57\pm0.04$\,mas/pixel.

\subsection{Background}\label{section:background}

Background measurements were obtained in five different low-background fields with F115W and F200W in both imaging and GR150C/R spectroscopy. Each measurement was compared with the prediction from the nominal background model at the date of the observation, all yielding an average background of $0.80\pm0.02$ relative to the nominal pre-flight background model. More details on background measurements for all instruments are reported in \citet{Rigby2023}.

\subsection{Scattered Light Issues}
\subsubsection{``Light saber"}

\begin{figure*}[!htbp]
    \centering
    \includegraphics[width=\linewidth]{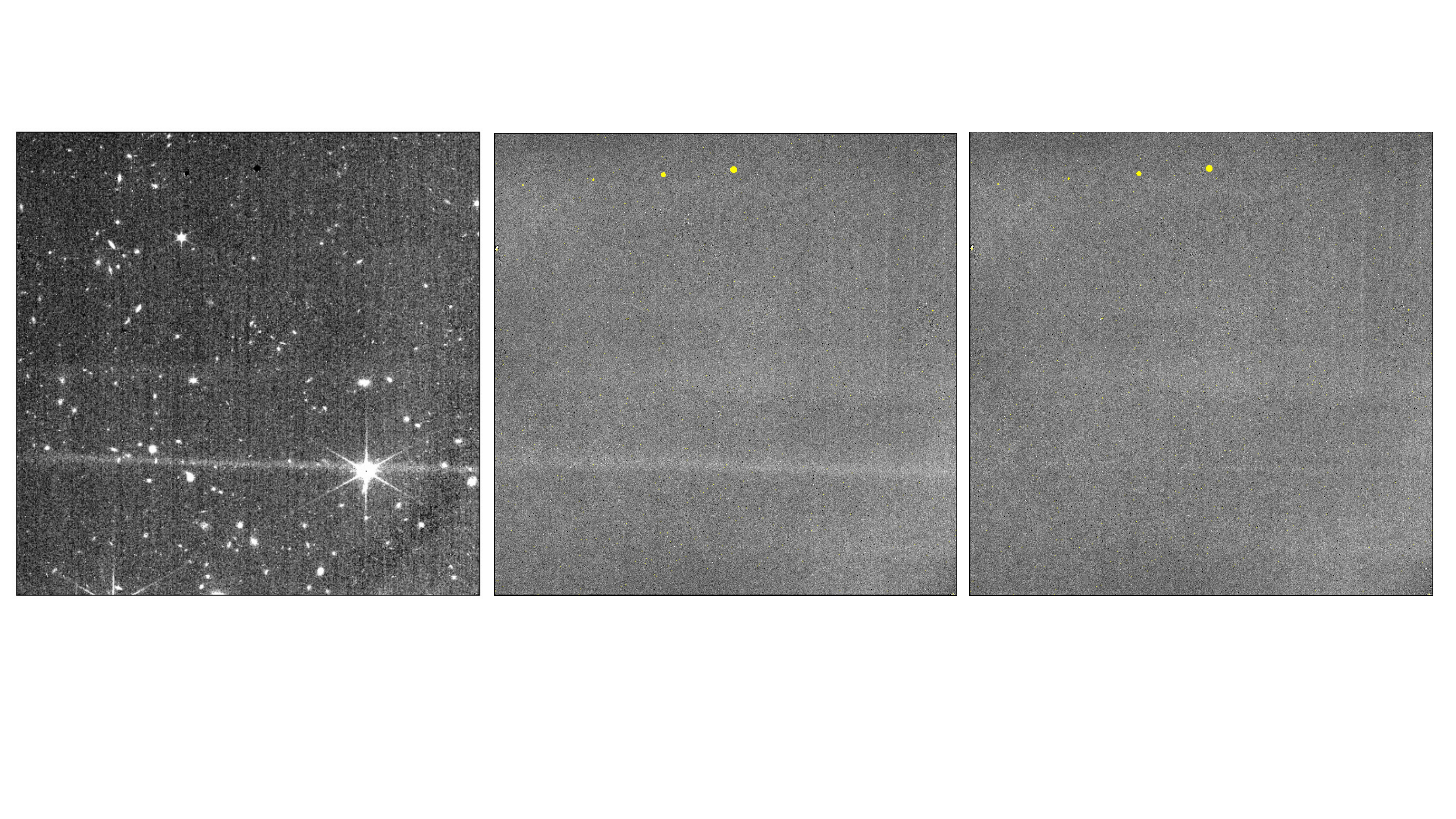}
    \caption{{\it Left}: F150W image stack showing a bright light saber horizontal feature due to a bright ($H\sim1$) star located within the out-of-field susceptibility region. {\it Middle}: F200W image of a typical lightsaber feature due to zodiacal background. {\it Right}: The same image as the middle panel but corrected by a two-dimensional model of the lightsaber. The circular features near the top left of the images are due to conical dimples directly engraved on the pick-off mirror acting as occulting spots for the coronagraphic mode of TFI \citep{Beaulieu2008}; those are permanent features of the NIRISS flatfield.} 
    \label{fig:LightSaber}
\end{figure*}

Imaging observations during commissioning revealed a nearly horizontal scattered light feature dubbed the ``light saber" whose root cause was identified, and very well modeled, as a rogue path originating from a susceptibility region 3.5\textdegree$\times$0.5\textdegree\, located approximately 22\textdegree~off-axis from the telescope FOV. The rogue path beam enters directly through the telescope entrance baffle, grazes off the wall following the NIRISS pickoff mirror, then makes a double reflection on two of the camera TMA mirrors before hitting the detector. The light saber is prominent when a very bright ($H_{vega}\sim1$) star is located anywhere within the susceptibility region but it is also detected with zodiacal light only, albeit at a much fainter level (see Figure~\ref{fig:LightSaber}). The light saber feature from integrated zodiacal light is well approximated by a two-dimensional, nearly Gaussian, function, hence can be easily subtracted out through a simple model (see right panel of Figure~\ref{fig:LightSaber}). The Astronomer Proposal Tool will also include a tool to warn users of a potentially bright lightsaber signal.  NIRCam is affected by similar scattered light features of different shapes, also originating from a rogue path \citep{Rieke2023}.

\subsubsection{Ghosts}

\begin{figure*}[!htbp]
    \centering
    \includegraphics[width=\linewidth]{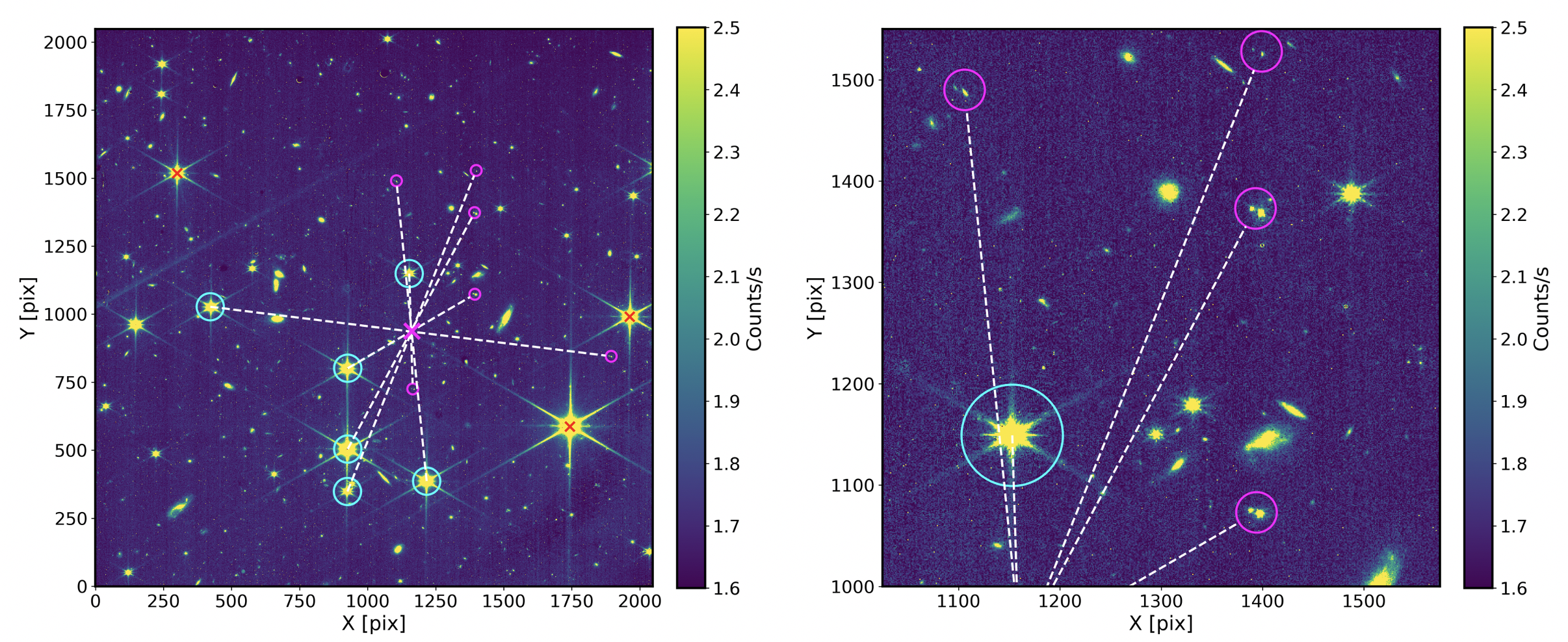}
    \caption{{\it Left}: F150W on-sky image of a field with several stars at different field positions (cyan circles) and their corresponding ghosts (magenta circles) connected by dashed white lines. The intersection of all lines (magenta cross) corresponds to the Ghost Axis Point. {\it Right}: Zoomed-in view to show the double nature of the ghosts for this particular filter. Figure from the \href{https://jwst-docs.stsci.edu/jwst-near-infrared-imager-and-slitless-spectrograph/niriss-performance/niriss-ghosts}{NIRISS Ghost} section of the JDox documentation.} 
    \label{fig:Ghosts}
\end{figure*}

NIRISS is affected by some image ghosts originating from internal reflections within the optical components of the dual wheel. Their relative intensity is field-dependent and varies between 0.1 and 4\%. A given filter ghost is characterized by a ghost axis point (GAP) corresponding to the intersection point of all lines between the source and the ghost for different field positions (see Figure~\ref{fig:Ghosts}). GAPs allow the user to predict ghost locations. Ghosts intensity and their GAPs were characterized during commissioning and previous ground-based tests campaigns.  They can be partly mitigated through dithering and most efficiently with the {\it MEDIUM} and {\it LARGE} dither patterns. AMI is not affected by ghosts since those fall outside its small FOV (80$\times$80 pixelx). No significant ghost affects the SOSS mode. More details on NIRISS ghosts are provided in the \href{https://jwst-docs.stsci.edu/jwst-near-infrared-imager-and-slitless-spectrograph/niriss-performance/niriss-ghosts}{NIRISS Ghost} section of the JDox documentation.

\subsubsection{Extra Diffraction spike}
\begin{figure}[!htbp]
    \centering
    \includegraphics[width=\linewidth]{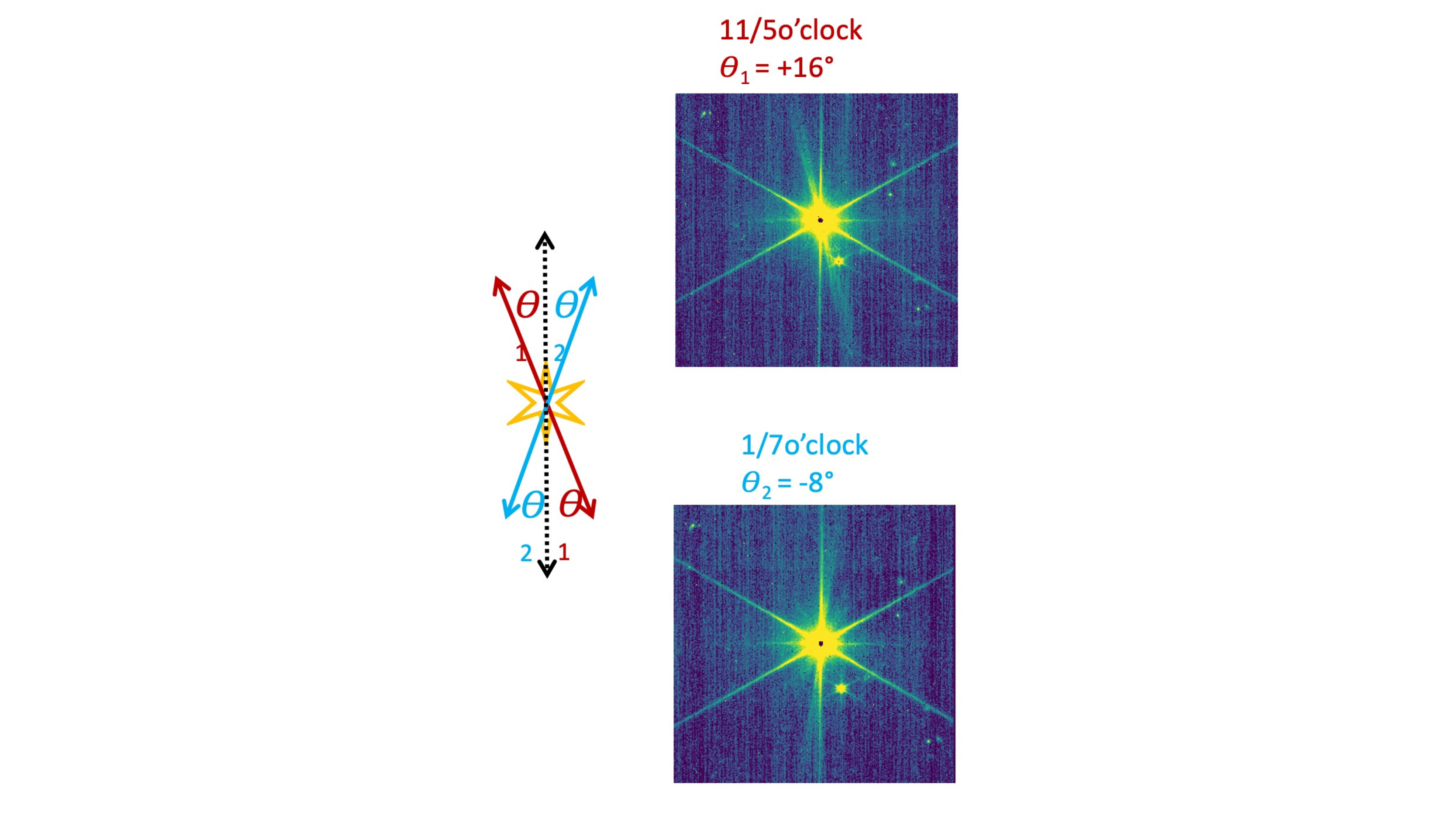}
    \caption{{\it Top}: Image of a bright star in F090W featuring an additional diffraction spike at a position angle (PA)  of +16\textdegree\, (11/5 o'clock). {\it Bottom}: same image but at a different $X\sim1400$ detector position, showing a similar diffraction spike but at a different PA of $-8$\textdegree\,(1/7 o'clock)} 
    \label{fig:Diffraction-spikes}
\end{figure}

    Images of bright stars show a diffraction-like spike that rotates with field position, mostly along the $X$ (V2) direction of the detector (see Figure~\ref{fig:Diffraction-spikes}). Its intensity, brightest in F090W and decreasing towards longer wavelength filters, is fainter ($\sim70$\%) than the regular diffraction spikes. In practice, this feature has a very small, hardly measurable, impact on the image quality.
    
    A similar field-dependent feature is also observed in the two FGS guiders. The most likely root cause of this effect for both instruments is small polishing errors within the TMAs associated with the diamond-turning process for manufacturing the FGS and NIRISS aluminum mirrors.    

\section{NIRISS Science Performance}

\subsection{Imaging and WFSS Sensitivity}
As described in previous sections, the measured throughput and image quality of NIRISS are significantly (10-30\%)  better than pre-flight predictions, particularly at short wavelengths, translating into improved sensitivity. While the notional 10$\sigma$ 10$^4$\,s sensitivity was not directly measured during commissioning with 10 individual 1000\,s exposures, one can make new predictions based on in-flight measurements. 

\begin{figure}[!htbp]
    \centering
    \includegraphics[width=\linewidth]{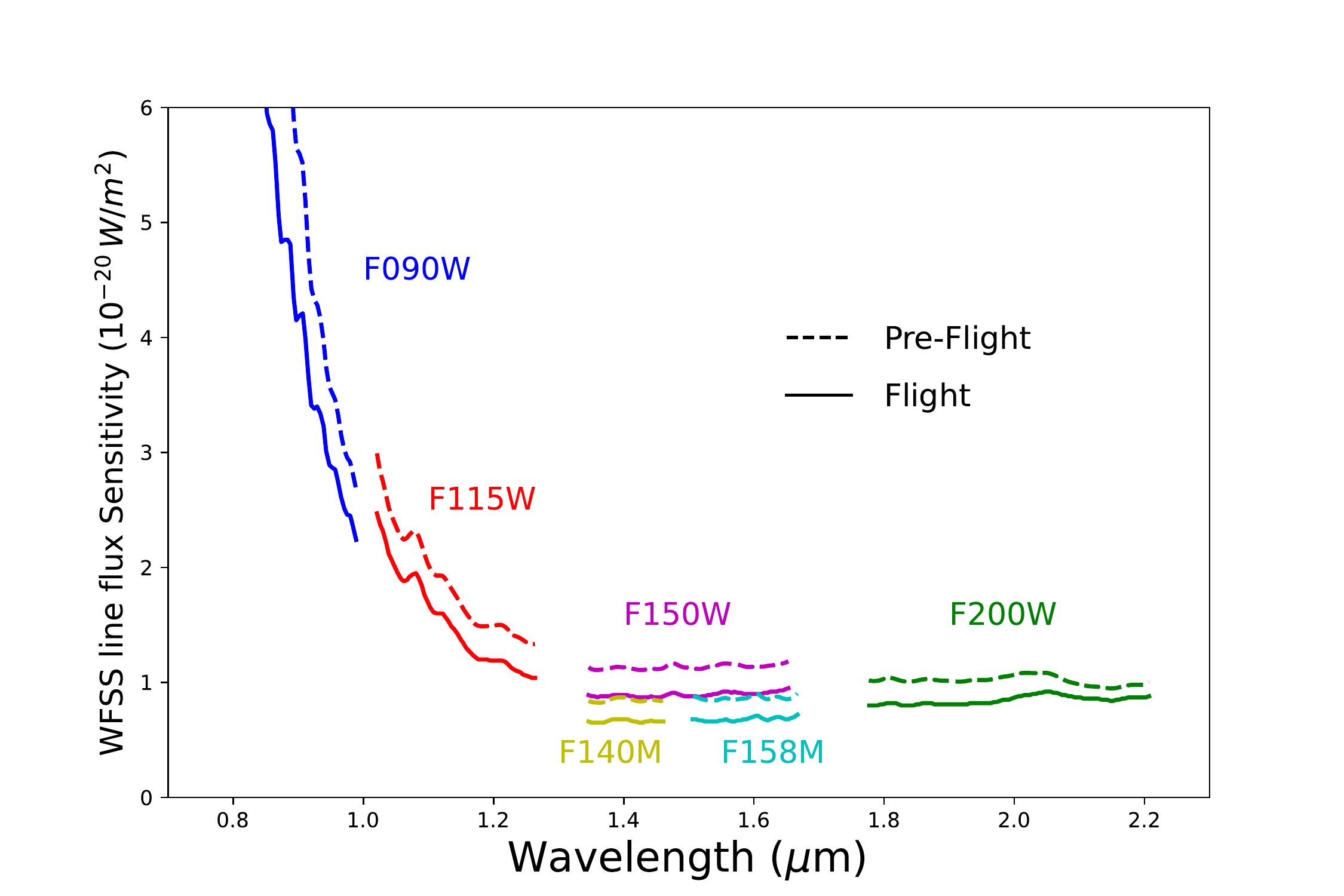}
    \caption{Comparison of pre- and in-flight line-flux sensitivity of the WFSS mode. The sensitivity improvement from pre-flight predictions varies between 20\% for F200W to 38\% for F090W.} 
    \label{fig:WFSS-sensitivity}
\end{figure}

Table~\ref{table:ImagingSensitivity} compares the pre-flight and in-flight continuum sensitivity predictions for all NIRISS filters. The new sensitivity assumes a zodiacal background model flux at a date of June 19, 2023 at RA=17h26m50.96s and Dec = -73\textdegree20\arcmin03.43\arcsec. The sensitivity improvement is typically $\sim$15\% at long wavelengths and between 24\% (F200W) and 44\% (F090W) in short-wavelength filters. As shown in Figure~\ref{fig:WFSS-sensitivity}, a similar improvement is observed for the WFSS line flux sensitivity, varying between 20\% in F200W and 38\% for F090W.

\begin{deluxetable*}{lccc}
\tablecaption{NIRISS Imaging sensitivity (10$\sigma\,10^4\rm s$)}
\tablehead{
\colhead{Filter} &  \colhead{ETC 1.0$^{\rm b}$} & \colhead{ETC 2.0$^{\rm b}$} & \colhead{$\Delta_{Sens}^{\rm c}$} \\
\colhead{} &  \colhead{(nJy)} & \colhead{(nJy)} & (\%)}
\startdata
F090W & 15.3 & 10.6 & +44\\
F115W & 13.0 & 9.5 &  +37\\
F140M & 17.6 & 12.9 & +36\\
F150W & 11.8 & 8.6 & +37 \\
F158M & 16.1 & 12.1 & +33\\
F200W & 10.2 & 8.2 & +24\\
F277W & 13.8 & 11.9 & +16\\
F356W & 14.5 & 12.6 & +15\\
F380M & 37.3 & 32.9 & +13\\
F430M & 51.4 & 46.0 & +12\\
F444M & 22.8 & 19.9 & +15\\
F480M & 63.4 & 56.5 & +12\\
\enddata
\tablecomments{$^{\rm a}$ Pre-flight ETC predictions. \\
$^{\rm b}$ In-flight ETC predictions. \\
$^{\rm c}$ Sensitivity improvement from pre-flight predictions ($\Delta_{Sens}=(ETC 1.0-ETC2.0)/ETC1.0)$ }
\label{table:ImagingSensitivity}
\end{deluxetable*}

\subsection{Time Series Observations}
The SOSS mode was tested and qualified through two time series observations (TSO): one 4-hr long sequence on an early-type star (BD+601753, A1V $J=9.6$) and a 6-hr long sequence on Hat-P-14, an F5V ($J=9.1$) star hosting a gas giant transiting exoplanet \citep{Torres2010} expected to yield a relatively small atmospheric signal at the level of 20-30 ppm. The latter was also observed by NIRCam \citep{Rieke2023, Schlawin2023} and NIRSpec \citep{Boker2023,Espinoza2023} to qualify their respective TSO modes. A detailed description of the NIRISS SOSS mode and flight performance is presented in \citet{Albert2023}. Here we present a high-level summary of the in-flight performance.

\subsubsection{BD+601753}

The TSO sequence on BD+601753 was used to qualify the stability and noise performance of the SOSS mode. 
The sequence comprised 876 consecutive 22-sec integrations (3 reads per integration) with maximal signal levels approximately halfway to saturation.  Except for a jump flux of $\sim$370 ppm halfway through the sequence (integration 413), the signal was very constant with a dispersion of 155 ppm within 5-10\% of the expected noise for such  white light observations. The jump flux is thought to be attributed to a so-called ``tilt event" whereby one of the primary segments of the telescope suddenly experiences a change in its tip/tilt configuration. A similar tilt event was also observed during the NIRCam TSO of Hat-P-14 \citep{Schlawin2023}. Thanks to the built-in weak lens in the GR700XD grism, a tilt event, or any other significant wavefront error variation of the primary mirror segments can be easily detected and accurately quantified with SOSS, either through the second derivative of the trace profile along the spatial direction or with principal component analysis (more details in \citet{Albert2023}). A detailed analysis of the TSO has shown that the jump flux is slightly chromatic due to the wavelength dependence of the diffraction-limited PSF in the spectral direction. 



\begin{figure}[!htbp]
    \centering
    \includegraphics[width=\linewidth]{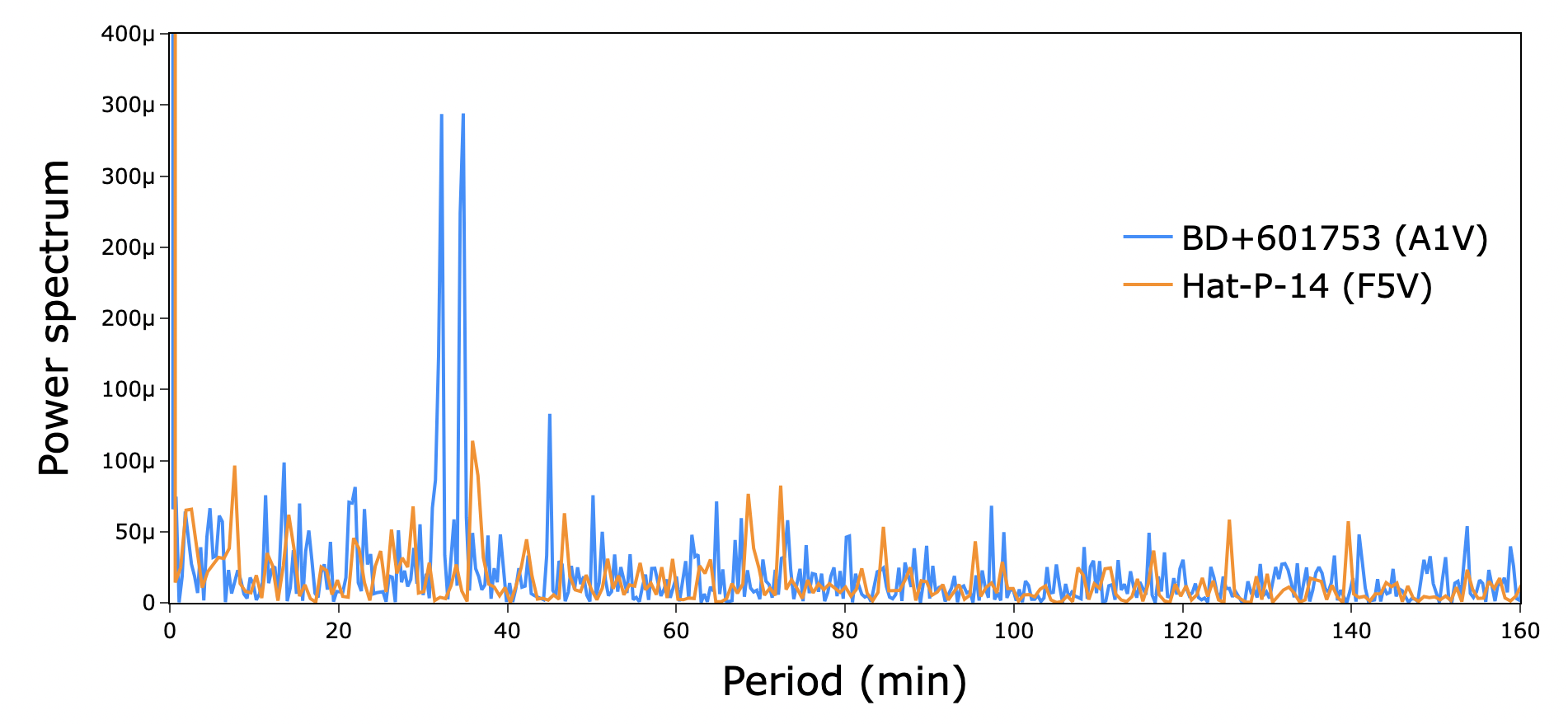}
    \caption{Power spectra of the TSOs of BD+6017653 and Hat-P-14 (corrected for the transit depth) \st{overlaid with the one sigma noise envelope}. The two different power spectra is suggestive of the stellar origin for these variations.} 
    \label{fig:SOSS-FFT}
\end{figure}

Aside from the tilt event, the power spectrum of the TSO (see Figure~\ref{fig:SOSS-FFT}) shows some clear coherent variability most likely of astrophysical origin (stellar pulsations) since a different power spectrum was observed on another star (Hat-P-14). Correcting for the tilt event and binning with time yields a noise varying as expected, i.e. as $1/\sqrt{N_{bin}}$ where $N_{bin}$ is the number integrations, down to $\sim$15 ppm with $N_{bin}$=200. 

Another evidence of the stellar origin of the observed variability is the fact that the continuum slope was  measured to be different in two stars. Indeed, the first hour on BD+601753 showed a significant slope of $\sim$-2 ppm/min while the whole 5-hr sequence was consistent with no variation (0.04$\pm$0.05 ppm/min). On the other hand, for HAT-P-14b, the first hour showed a continuum slope of -0.4 ppm/min and 0.75$\pm$0.16 ppm/min for the whole sequence. 

\subsubsection{Hat-P-14}

A transit event of HAT-P-14b was captured by SOSS in June 2022 during a 6-hr sequence consisting of 572 consecutive integrations with Ngroup$=6$. The detailed analysis of this sequence in described elsewhere \citep{Albert2023}. 
Like BD+601753, a tilt event was detected at the beginning of the sequence (integration 60, $\sim\,-3$ hours before mid-transit) but a much smaller one with a very minor impact on the white light curve (WLC). Similarly to BD+601753, the power spectrum of the TSO of Hat-P-14, corrected for the transit event, also show some coherent, albeit different, frequencies, suggesting that Hat-P-14 is inherently variable. Overall, the best WLC fit model is obtained with a Gaussian process regression. The resulting transit spectra for both orders is shown in Figure 27 of \citet{Albert2023}. Unlike the flat spectrum observed between 2.8 and 4.8 $\mu$m with the NIRSpec G395 mode \citep{Espinoza2023}, the SOSS spectrum binned to R=50 shows some significant deviations from flatness at the level of $\sim$100 ppm with median transit depth uncertainties of 50 ppm. 

The integration time of this TSO was deliberately chosen to reach near saturation around the blaze peak (1.3 $\mu$m) to investigate the noise behavior, i.e. the flattening of the signal-to-noise (SNR) for signal levels close to saturation.  It was found that the SNR is flattening for signals higher than $\sim$35000 ADUs (56000 e-) i.e. 3/4 of the detector well dept ($\sim$72000 e-). While this requires further investigation with more data, it is recommended to set an appropriate number of reads (Ngroups) to remain below $\sim$50000 e- at the end of the integration to avoid potential SNR saturation.

In summary, aside from tilt events, inherent stellar variability and SNR saturation beyond a signal of $\sim$50000 e-, no significant source of systematic noise of instrumental origin could be detected in both TSOs. This demonstrates the power of SOSS not only to deliver high-accuracy spectrophotometry for its main intended application, transit/eclipse spectroscopy of relatively bright stars, but also for other applications like asteroseismology. SOSS observations allows disentangling genuine variable signals of astrophysical origin from systematic flux variations associated with wavefront error changes of the telescope. 


\subsection{Companion Detection with AMI and KPI}

The point source detection capability of AMI was tested on the young quadruple system AB Dor featuring two binary pairs: AB Dor A/C and AB Dor Ba/Bb separated by  9\arcsec \citep{Close2005, Guirado2006, Azulay2017}. AB Dor A/C  is a low-mass companion near the hydrogen burning limit at a current separation of $\sim$0.3 arcsec with a contrast of 4.5\,mag, well within the AMI detection sensitivity. 

\begin{figure}[!htbp]
    \centering
    \includegraphics[width=\linewidth]{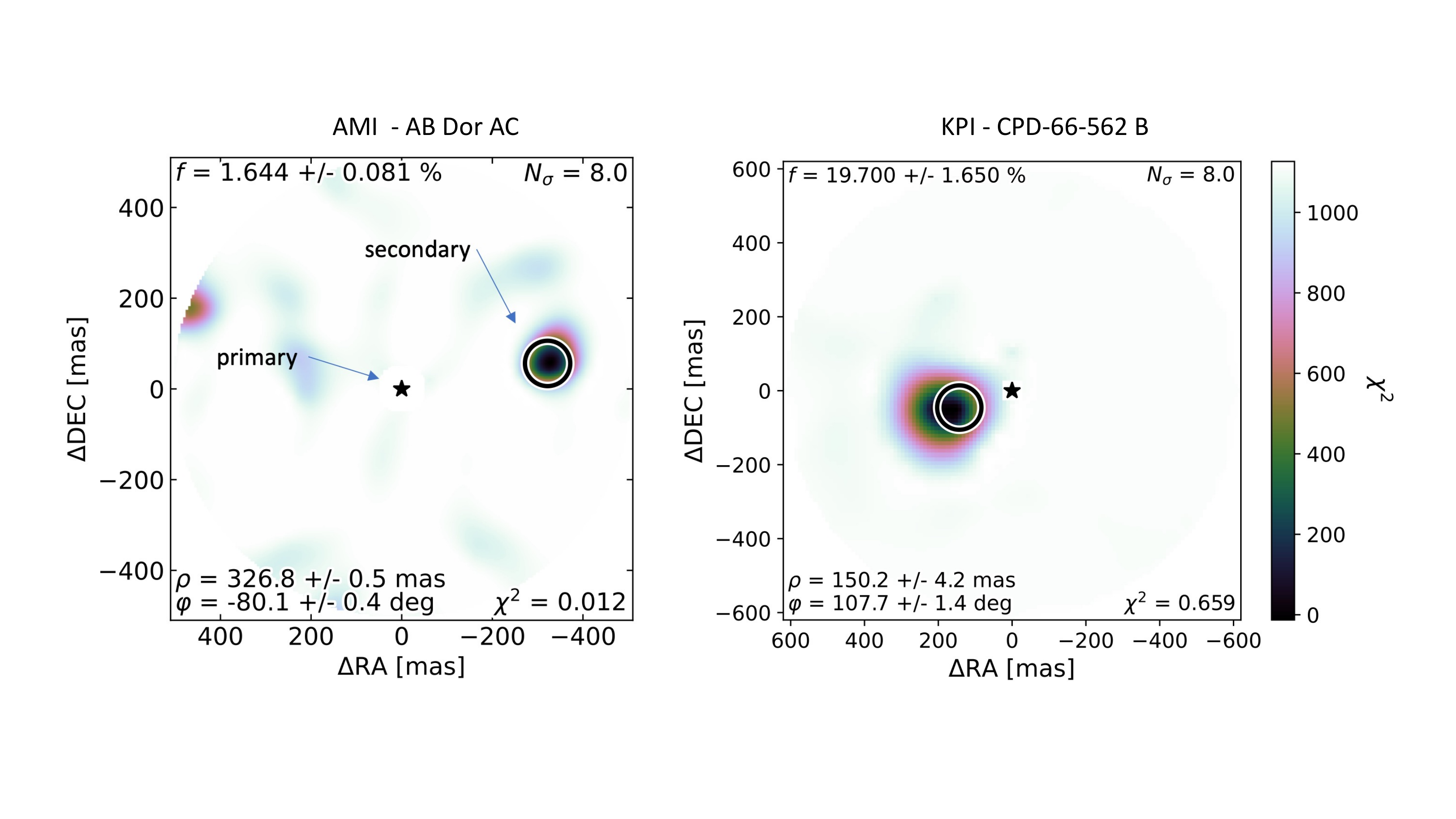}
    \caption{{\it Left}: $\chi^2$ map of the best fit model of the observed closure phase and visibility for the star AB Dor, known to host a companion $\sim$350 mas and 4.5 magnitudes fainter. The companion is unambiguously detected with the expected contrast $f$, position angle $\phi$ and a separation $\rho$ within 3$\sigma$ of the prediction. {\it Right}: Similar $\chi^2$ map with the KPI point source detection algorithm applied to the star CPD-66-562 showing a $\sim$1:5 contrast companion detected at $\sim$150 mas. }    
    \label{fig:AMI-KPI}
\end{figure}

AB Dor A was observed at two epochs during commissioning (PID 1093) in all three AMI filters (F380M, F430M, F480M) along with the star HD37093 of similar spectral type (K5III) used as a calibrator. Both targets were positioned on the 80$\times$80 AMI subarray with the standard target acquisition procedure accurate to $\sim1/20^{th}$ of a pixel. Figure~\ref{fig:AMI-KPI} (left panel) shows the $\chi^2$ map of the point source model fitting both the closure phases and visibility amplitudes computed with fouriever\footnote{\url{https://github.com/kammerje/fouriever}}\citep{Sivaramakrishnan2023}. AB Dor C was unambiguously detected with a contrast of 4.51$\pm$0.05, a separation $\rho$ and position angle $\phi$ of 326.8$\pm$0.5 mas and -80.1$\pm$.4\textdegree, respectively, in reasonable agreement with the measured $L$-band contrast of 4.2$\pm$0.3\,mag \citep{Azulay2017} and separation $\rho=369.4\pm$15 mas and position angle $\phi=$-78.3$\pm$20\textdegree\, predicted from the dynamical model (private communication). The depth of the AB Dor observations ($\sim10^8$\,photons) enables a 3$\sigma$ companion detection limit of $\sim$7.5\,mag near $\sim$200 mas \citep{Sivaramakrishnan2023}, between $0.5-1$\,mag worse than theoretical predictions. 

The KPI point detection capability was also tested using four stars with unknown binary statuses, all acting as calibrators with one another. As shown in Figure~\ref{fig:AMI-KPI} (right panel), one target, CPD-66-562, was serendipitously discovered to have a companion at 150 mas with a contrast of 1.8\,mag. Another companion with a contrast of $\sim$5.6 mag was also detected around 2MASSJ062802.01-663738 at a separation of $\sim$240\,mas. From these observations, one infers a 5-$\sigma$ companion detection limit of $\sim$6.5 mag at $\sim$200 mas and $\sim$7 mag at $\sim$400 mas  \citep{Kammerer2023}. Both KPI and AMI are complementary with one another, the former better optimized for faint targets with companions beyond $\sim$325 mas while AMI should be preferred to search for relatively faint companions inward of $\sim$325 mas around bright targets. 

\section{Summary}
We have presented an overview of the NIRISS instrument, a description of its design, and four observing modes: imaging, wide-field slitless spectroscopy, single-object slitless spectroscopy, and aperture masking interferometry. In-flight data have qualified all four  modes with performance exceeding pre-flight predictions. NIRISS shows a significantly better response at short wavelengths where most of its science programs are concentrated. More specifically, the imaging and WFSS modes show improved sensitivities between 10 and 40\% and the SOSS mode has an improved throughput of $\sim$60\% in the second order. Time series observations with SOSS show very stable spectro-photometric performance within 10-20\% of the expected noise. Finally, the first companion detection through space-based aperture masking interferometry was demonstrated with NIRISS.

\acknowledgments

 We thank the anonymous referee for constructive comments and suggestions that improved the quality of the paper. NIRISS and FGS are the fruit of more than two decades of teamwork by hundreds of persons of all origins and diversity across Canada at the CSA, industry (in particular Honeywell Aerospace), universities, the National Research Council Canada (NRC) in collaboration on the JWST project with thousands of others from NASA, ESA and US/European institutions. Special thanks to Paul Kuzmenko and Erich Bach for their dedication in the manufacture and timely delivery of the NIRISS grisms. The Canadian contribution to JWST would not have been possible without the scientific vision and leadership of Simon Lilly. We are grateful to the late Phil Sabelhaus, former Project Manager of JWST, Bill Ochs and the JWST Science Working Group for their enabling decisions at critical times during the Project that contributed to the successful and timely transition between TFI and NIRISS. The Canadian contribution to JWST results from an unprecedented, and several times renewed, financial contribution from CSA. We also acknowledge financial support from the Université de Montréal, NRC, the Natural Science and Engineering Research Council of Canada, {\it Développement Économique Canada}, the {\it Fonds de Recherche Québécois - Nature, Technologie et Santé} and the Trottier Family Foundation. D.J. is supported by NRC Canada and by an NSERC Discovery Grant. Space Telescope Science Institute is operated by the Association of Universities for Research in Astronomy, Inc., under NASA contract NAS 5-03127 for JWST. Finally, we thank the general public for their general support and communicative enthusiasm for this historic space mission. 

\bibliography{NIRISS-overview.bib}{}
\bibliographystyle{aasjournal.bst}

\end{document}